\documentclass[journal]{IEEEtran}
\usepackage[utf8]{inputenc}
\usepackage{xcolor}
\usepackage[ruled,norelsize]{algorithm2e}
\usepackage{tikz}
\usepackage{amsmath}
\usepackage{amsfonts}
\usepackage{amssymb}
\usepackage{mathtools}
\usepackage{amsthm}
\usepackage{bm} 
\usepackage{tikz}
\usepackage{subcaption}
\usepackage{multirow}
\usepackage{cite}
\usetikzlibrary{arrows.meta}

\DeclarePairedDelimiter\floor{\lfloor}{\rfloor}

\makeatletter
\newcommand{\removelatexerror}{\let\@latex@error\@gobble}
\makeatother

\newtheorem{lem}{Lemma}
\newtheorem{prop}{Proposition}
\newtheorem{corollary}{Corollary}

\newcommand{\cP}{\mathcal{P}}
\newcommand{\cS}{\mathcal{S}}
\newcommand{\cT}{\mathcal{T}}
\newcommand{\cU}{\mathcal{U}}

\newcommand{\bp}{\mathbf{p}}

\newcommand{\lt}{\log_2}
\newcommand{\wl}{w_\ell}

\newcommand{\bbN}{\mathbb N}
\newcommand{\jj}{{\sf j}}

\begin{document}

\title{Binary-Tree Encoding for Uniform Binary Sources in Index Modulation Systems}

\author{Justin~P.~Coon,
Mihai-Alin~Badiu,
Ye~Liu,
Ferhat~Yarkin,
Shuping~Dang
\thanks{J. P. Coon, M.-A. Badiu, Y. Liu, and F. Yarkin are with the Department of Engineering Science, University of Oxford, Parks Road, Oxford, OX1 3PJ, UK, (Email: justin.coon@eng.ox.ac.uk).}
\thanks{M.-A. Badiu is also with the Department of Electronic Systems, Aalborg University, Fredrik Bajers Vej 7, Aalborg, 9220, Denmark.}
\thanks{S. Dang is with Computer, Electrical and Mathematical Science and Engineering Division, King Abdullah University of Science and Technology (KAUST), Thuwal 23955-6900, Kingdom of Saudi Arabia.}
\thanks{This work was supported by EPSRC grant numbers EP/R511742/1 and EP/N002350/1.}
}

\maketitle

\begin{abstract}
The problem of designing bit-to-pattern mappings and power allocation schemes for orthogonal frequency-division multiplexing (OFDM) systems that employ subcarrier index modulation (IM) is considered.  We assume the binary source conveys a stream of independent, uniformly distributed bits to the pattern mapper, which introduces a constraint on the pattern transmission probability distribution that can be quantified using a binary tree formalism.  Under this constraint, we undertake the task of maximizing the achievable rate subject to the availability of channel knowledge at the transmitter.  The optimization variables are the pattern probability distribution (i.e., the bit-to-pattern mapping) and the transmit powers allocated to active subcarriers.  To solve the problem, we first consider the relaxed problem where pattern probabilities are allowed to take any values in the interval $[0,1]$ subject to a sum probability constraint.  We develop (approximately) optimal solutions to the relaxed problem by using new bounds and asymptotic results, and then use a novel heuristic algorithm to project the relaxed solution onto a point in the feasible set of the constrained problem.  Numerical analysis shows that this approach is capable of achieving the maximum mutual information for the relaxed problem in low and high-SNR regimes and offers noticeable benefits in terms of achievable rate relative to a conventional OFDM-IM benchmark.
\end{abstract}

\section{Introduction}
As a subclass of permutation modulation \cite{Slepian1965}, index modulation (IM) has recently attracted significant interest \cite{Basar2017,Ishikawa2018} due to its feature of ``achieving more by doing less''. The central idea of IM lies in the observation that, in addition to encoding information in a signal, one can encode information in the order in which a signal is conveyed in a given domain.  The idea of encoding information using permutations or combinations has been applied in several contexts.  For example, by using different transmit antennas and channel uniqueness, permutation modulation has been employed in the spatial domain in the form of so-called spatial modulation \cite{Mesleh2008,DiRenzo2014}.  Similar ideas have been applied to the medium/channel domain by manipulating the radiation patterns of antennas~\cite{Khandani2013,Basar2017STCM}.  Permutation modulation has also been used in the subcarrier index domain in multicarrier systems, such as orthogonal frequency-division multiplexing (OFDM).  This approach is commonly referred to as subcarrier-IM or simply IM \cite{Tsonev2011,Basar2013}.  Finally, the use of permutation methods in conjunction with different modes in orbital angular momentum transmissions has been studied~\cite{Willner2016,Yang2018}.

To facilitate the use of combinatorial patterns for encoding, a codebook for the mapping between patterns and the source messages (bit sequences) must be specified. Many existing works that study permutation modulation in digital communication systems assume that the number of possible patterns is a power of two \cite{Mesleh2008,Yang2008,Maleki2013}. However, such an assumption limits the applicability of permutation modulation, e.g., conventional spatial modulation (with a single active antenna in each transmission period) is only applicable when the number of antennas at the transmitter is a power of two.

Another typical approach that has been studied is to assume that only a subset of all possible patterns contains valid patterns, and the size of the subset is a power of two~\cite{Younis2010,Wang2012,Basar2013,Dang2018}.  However, this approach is not able to utilize the full potential of permutation modulation in terms of data rate, because a certain number of possible permutations that could have been used to carry information are neglected~\cite{Liu2019}. The study detailed in~\cite{Wen2016} considers the possibility of using all permutation patterns with uniform probability, but no treatment of how to realize the uniform probability distribution in digital communication systems is given in that work. 


To address these issues related to the mapping of source bit sequences to permutation patterns, a few recent contributions have focused on the adaptation of binary Huffman coding~\cite{Huffman1952} for permutation/index codebook design~\cite{Siddiq2016,Wang2017,Hu2018,Kadir2018,Yang2018,Liu2019}.  Here, a bijective mapping between information bit sequences and the permutation/index patterns is constructed with the aid of a full binary tree; patterns are associated with leaves in the tree, and corresponding bit sequences are defined according to a labeling rule (used in the Huffman algorithm) pertaining to the respective paths from each leaf to the root.  Importantly, in contrast to conventional application scenarios for source compression where the source symbol distribution is known a priori, the probability distribution of the patterns observed during transmission in permutation modulation systems \emph{is dependent upon the binary source}~\cite{Liu2019}. In this sense, the Huffman mapping is applied in permutation modulation schemes in a reversed manner.  We adopt the term \emph{binary-tree encoding} rather than \emph{Huffman coding} for the bit-to-pattern mapping operation for the remainder of this paper in order to highlight this subtle, but important difference.

Binary-tree encoding for permutation modulation schemes enables one to choose the probability distribution of the permutation patterns to achieve certain design criteria, e.g., achievable rate maximization~\cite{Wang2017,Kadir2018,Yang2018,Liu2019} or symbol-error rate (SER) minimization~\cite{Wang2017,Kadir2018}. However, existing works along this direction fall short in a number of ways.  For example, the support of the (random) patterns, when constrained by full binary tree structures, is discrete. As a result, optimization problems for maximizing achievable rates or minimizing SERs are of mixed-integer forms, and an exhaustive search over all admissible probability distributions may be required to find the global optimum.  However, the number of admissible distributions has not been characterized in the literature, and thus the complexity of exhaustive searching is not well understood.  A common way to reduce optimization complexity that has been treated in the literature is to relax the full-binary-tree constraint on the pattern probability distribution~\cite{Wang2017,Yang2018}. However, the problem of how to project the relaxed probability distribution to a feasible distribution that satisfies the full-binary-tree constraint remains open.  An alternative strategy that has received attention recently has been to focus on high and low signal-to-noise ratio (SNR) regimes.  For the limited case of single-active-antenna spatial modulation, analytic forms of the asymptotically optimal probability distributions for the permutation patterns were reported in~\cite{Wang2017}.  A generalization that activates multiple resources per channel use, which is the scenario of interest for multicarrier communication systems such as OFDM-IM~\cite{Basar2013}, is desirable.

In this work, we study the subclass of permutation modulation where $K$ out of $N$ resources are active during each channel use. We concentrate our investigation on OFDM-IM systems~\cite{Basar2013}, because OFDM-IM is a primary user of the permutation modulation subclass that we study, and any results obtained for full binary trees would be directly applicable to other permutation modulation schemes.  Our main goal is to optimize the bit-to-pattern mapping operation and transmit power allocation strategy for achievable rate maximization when channel state information is available at the transmitter. We make the following contributions.
\begin{enumerate}
    \item We give a complete and rigorous formulation of the bit-to-pattern mapping problem using the formalism of full binary trees, which covers all admissible pattern probability distributions given a uniform binary source.  To this end, we report a new method to generate a reduced set of these trees and establish bounds on the number of trees in this set, which have not been reported in the mathematics or engineering literature to the best of our knowledge.
    \item We formulate a relaxation of the achievable rate optimization problem with pattern probabilities and transmit powers as the optimization variables and give a number of analytic bounds and high/low-SNR asymptotic results that can be used to (approximately) solve the problem.
    \item We propose an efficient, heuristic algorithm that projects a relaxed pattern probability distribution onto the feasible set of distributions that obey the full binary tree constraints, and demonstrate that this method yields an achievable rate that is superior to a conventional OFDM-IM benchmark.
\end{enumerate}

The rest of the paper is organized as follows.  In Section~\ref{sec:model}, the basic OFDM-IM model is described, with emphasis being placed on the binary-tree encoding operation.  Section~\ref{sec:trees} explores the fundamental properties of binary trees; in this section details of the new tree construction algorithm are provided along with proof of completeness and bounds on the number of trees of a given size are reported.  In Section~\ref{sec:rate-relaxed}, a relaxation of the achievable rate optimization problem is explained, and several analytic bounds and asymptotic results related to this problem are given.  The fully constrained optimization problem is treated in Section~\ref{sec:rate-constrained}, where the aforementioned heuristic projection algorithm is outlined.  A numerical analysis of all results reported in the paper are included in Section~\ref{sec:numerics}, and conclusions are drawn in Section~\ref{sec:conclusions}.

\section{System Model}
\label{sec:model}

\subsection{Binary-Tree Encoding}
Consider a binary sequence $\{B_n\}_{n\in\bbN}$, which is conveyed from a maximum entropy source to an OFDM-IM encoder.  The maximum entropy property of the source implies the sequence elements $B_n\in\{0,1\}$ are independent, uniformly distributed random variables.  The encoder partitions\footnote{The exact detail of how this partitioning is accomplished is beyond the scope of this paper.} the sequence $\{B_n\}$ into two subsequences $\{B_{n_k}\}$ and $\{B_{n_\ell}\}$, where $k,\ell \in \bbN$ and $k \neq \ell$.  One subsequence (say, $\{B_{n_k}\}$) is mapped to a sequence of $M$-ary complex-valued constellation symbols.  For example, if 16-QAM is employed, $M=16$, and each group of $m = \lt M = 4$ bits in $\{B_{n_k}\}$ is mapped to a QAM symbol.  The other subsequence is used to assign the $M$-ary symbols to subcarriers in preparation for transmission.  In the IM system considered in this paper, we assume each OFDM symbol vector is comprised of $G$ groups of $N$ subcarriers, and $K \leq N$ subcarriers in each group are active, while the remaining $N-K$ subcarriers are nulled\footnote{We will focus on the case where the inequality is strict, since $K=N$ corresponds to a conventional OFDM system.}.  In keeping with convention, we use the term \emph{subcarrier activation pattern (SAP)} to denote a pattern of $K$ active subcarriers (out of $N$).  

We are interested in system designs that maximize the achievable rate of OFDM-IM; hence, we consider bit-to-SAP mapping strategies that cover the full set of available SAPs.  Since there are $\binom{N}{K}$ SAPs, it is generally not possible to construct a fixed-length bit-to-SAP mapping scheme that satisfies this condition.  For example, with $N=4$ and $K=2$, six SAPs are available.  By using a fixed-length mapping scheme, it would be possible to map two bits to one of four SAPs, leaving two SAPs unused.  

To overcome this issue, we employ a variable-length scheme based on \emph{full binary trees}.  A tree is a full binary tree if every node other than the leaf nodes has exactly two children.  Every full binary tree comprised of $v$ internal nodes has $v+1$ leaves.  Considering the total set of $v$-node full binary trees\footnote{For the rest of the paper, unless explicitly stated otherwise, a $v$-node full binary tree is one with $v$ internal nodes.}, the maximum depth of a tree in the set ranges from $\floor{\lt v} + 1$ to $v$.

\begin{figure}[t]
    \centering
    \includegraphics[angle=0,width=7cm,clip=false]
    {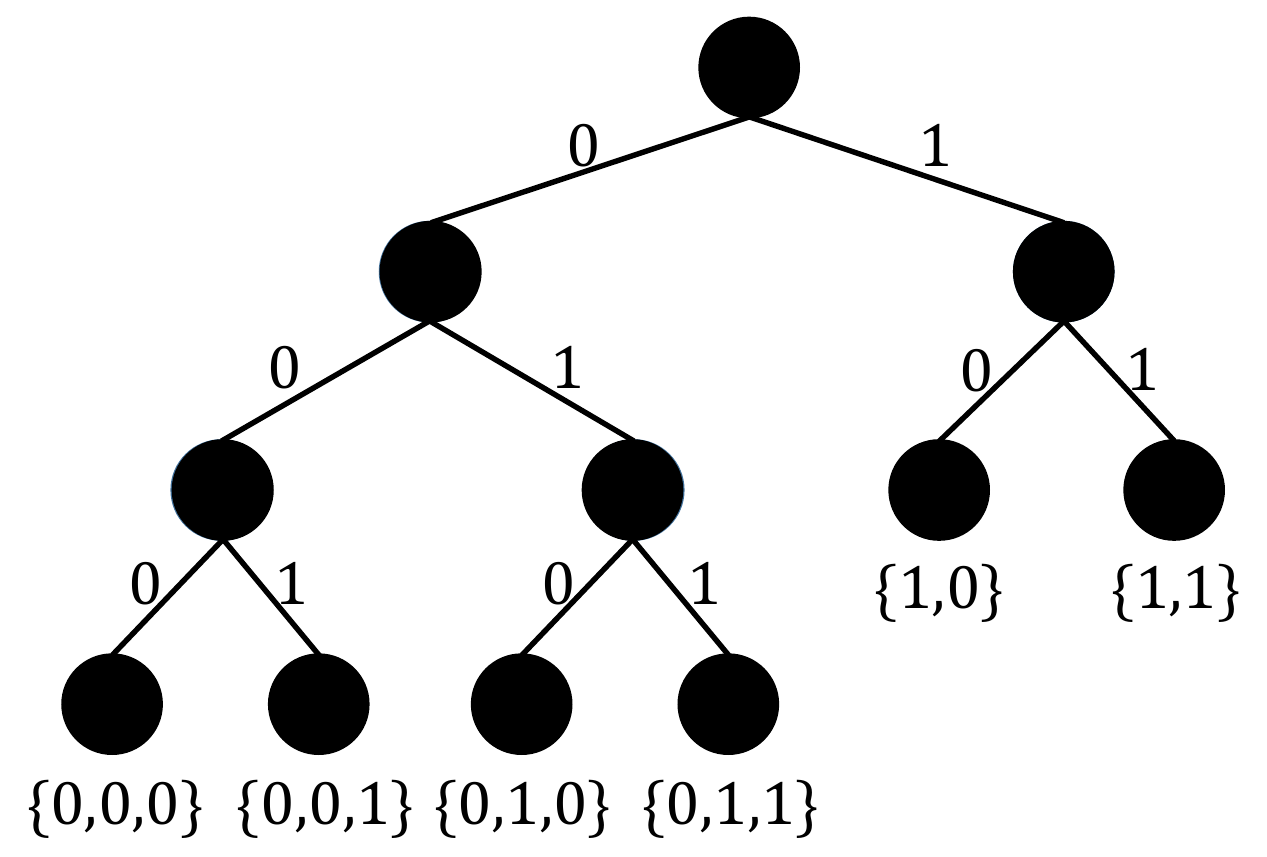}
    \caption{A possible bit-to-SAP mapping when $N=4$ and $K=2$. Each edge is labeled with zero or one.  Each SAP is associated with a unique leaf.  For each leaf, the set below is the bit sequence mapped to that leaf.}
    \label{fig:tree}
\end{figure}

It is well known that one can map symbols from a source alphabet to uniquely and instantaneously decodable bit sequences using full binary trees.  Indeed, this method is employed in the celebrated Huffman source coding algorithm.  For the IM system considered herein, we apply a \emph{reverse mapping} approach, which entails the use of a chosen binary tree to map source bit sequences to SAPs.  Each edge of the tree is labeled with a zero or a one, and the tree is constructed such that it has $\binom{N}{K}$ leaves.  Each SAP in the set of $\binom{N}{K}$ admissible patterns is associated with a leaf.  The bit-to-SAP mappings are determined by tracing the unique path from the root node to each leaf, recording the bit labels for each edge in order along the way.  Fig.~\ref{fig:tree} provides an illustration of this procedure for the example of $N=4$ and $K=2$.

Similar to Huffman source coding, the use of full binary trees to develop a bit-to-SAP mapping rule ensures each mapping is unique and instantaneously \emph{encodable}.  Uniqueness results from the binary tree structure.  Instantaneous encodability simply means that the encoder can map bit sequences to SAPs using the minimum amount of information.  To illustrate this point, we can again turn to Fig.~\ref{fig:tree}.  Suppose the SAP bit sequence is $\{0,0,1,1,0,1,1\}$.  Reading left to right and referring to Fig.~\ref{fig:tree}, we see that the encoder only needs to read the first three bits to map them to the SAP associated with the second leaf.  The encoder would then read $\{1,0\}$, which also yields a valid SAP (the fifth leaf), and so on.  In this example, it is clear that the encoder does not need to interpret long sequences of bits in order to decide upon the correct mapping relating to the first few bits.  

It is also important to note that \emph{every possible} two and three-bit sequence is accounted for in the mapping shown in Fig.~\ref{fig:tree}.  This property extends to mappings based on other values of $N$ and $K$. For a maximum entropy bit source, each subsequence consisting of $q$ bits will appear with probability $1/2^q$.  Thus, we immediately deduce that an SAP associated with a leaf node at level $q$ below the root will be transmitted with probability $1/2^q$.  \emph{This feature of binary-tree encoding imposes a constraint on the system, which much of the literature published on this topic to date has largely ignored.}  In this work, we will exploit the structure imposed by this constraint to develop efficient optimization procedures for OFDM-IM systems.

\subsection{OFDM Model}
Once bit-to-symbol mappings (both constellation and SAP) have been completed, each length-$GN$ OFDM symbol vector is processed with a $GN$-point inverse discrete Fourier transform (DFT) and a cyclic prefix of adequate length (to mitigate the effects of channel dispersion) is appended to each time-domain symbol array prior to filtering, up-conversion, and transmission.  At the receiver, the received signal is down-converted, filtered, and sampled.  The cyclic prefix is then removed from each received baseband symbol vector before processing with a DFT.  It is well-known that this sequence of procedures converts the dispersive channel into a parallel channel, and the signal on each subcarrier is (ideally) free of interference from other subcarriers.

We now formalize the OFDM model.  Define $C \coloneqq \binom{N}{K}$. We can uniquely associate each SAP with an index in the set $\mathcal{U}\coloneqq\{1,\ldots,C\}$. For each $i\in\mathcal{U}$, denote by $\mathcal{S}_i\subseteq\{1,\ldots,N\}$ the set of indices of the $K$ subcarriers that are active under pattern $i$, where equality holds when $K=N$ (which corresponds to a conventional OFDM system). The index symbol $U$ is randomly distributed over $\mathcal{U}$ with probabilities $p_i\coloneqq \mathbb{P}(U=i)$, $i\in\mathcal{U}$. The channel input-output relationship for subcarrier $l\in \{1,\ldots,N\}$ conditioned on the SAP can be written as
\begin{equation}
Y_l\mid U=i\text{ equals }
\begin{cases}
\sqrt{g_l}e^{\jj\theta_l} X_l + Z_l, &\text{if }l\in\mathcal{S}_i, \\
Z_l, &\text{if }l\notin\mathcal{S}_i,
\end{cases}
\end{equation}
where $\sqrt{g_l}e^{\jj\theta_l}$ is the complex channel coefficient for subcarrier $l$ (with $\jj = \sqrt{-1}$); the input symbols $\{X_l\}$ are zero mean and independent over the subcarriers with $\rho_{li}$ being the transmit power on subcarrier $l$ for index $i$; the noise is independent over the subcarriers with $Z_l\sim\mathcal{CN}(0,\sigma^2)$.  Throughout this paper, we will assume the channel gains $\{g_l\}$ are known at the transmitter.  We will return to this model in the context of mutual information optimization in Sections~\ref{sec:rate-relaxed} and~\ref{sec:rate-constrained}.

\section{Full Binary Trees}
\label{sec:trees}
One of the goals of this work is to develop a method of computing the bit-to-SAP mapping that maximizes the achievable rate of an OFDM-IM system.  This is equivalent to determining the full binary tree that defines the optimal mapping.  To achieve this aim, we will need a method of considering all full binary trees of a given size as well as all SAP-to-leaf assignments.  At first glance, this is a complicated problem.  The number of $v$-node trees is given by the Catalan number
\begin{equation*}
        c_v = \frac{1}{v+1}\binom{2v}{v} \sim \frac{4^v}{v^{3/2} \sqrt{\pi}}
\end{equation*}
and the number of SAP-to-leaf assignments is $(v+1)!$.  However, it is possible to significantly simplify the problem by making use of symmetry.  The important aspect of the mapping is not in the exact tree that is chosen, but rather in the level of the leaf node that a given SAP is assigned to.  As noted in Section~\ref{sec:model}, an SAP assigned to a leaf at level $q$ has probability $1/2^q$ of being transmitted.  We can transpose leaf nodes at a given level in any way we wish and still achieve the same SAP probability distribution.  This reasoning leads us to consider a smaller set of trees, which we call the \emph{reduced set of $v$-node full binary trees} $\cT_v$.  Each tree in this set actually corresponds to an automorphism group of the complete set.  Moreover, consider a given tree $t\in\cT_v$ and denote the number of leaves at level $q$ by $n_q$.  Due to the symmetry stated above, the number of ways of assigning $v+1$ objects (i.e., SAPs, where $v+1 = C$) to the leaf nodes such that we attain a unique probability distribution is
\begin{equation}
  \binom{v+1}{n_1,n_2,\ldots,n_v} = \frac{(v+1)!}{n_1!n_2!\cdots n_v!}
\end{equation}
which can be considerably smaller than the total $(v+1)!$ permutations.  We now give preliminary results on the construction and enumeration of the set $\cT_v$, which will be useful in determining systematic optimization procedures and analyzing computational complexity.

\subsection{Construction}
In order to choose the best tree for encoding, we require a method of constructing all trees in $\cT_v$.  The approach we propose is outlined in Algorithm~\ref{alg:badiu}, which is valid for $v\geq 2$.  The initial set $\cT_1 = \{\tau\}$ consists of the single full binary tree $\tau$ with one root and two leaves (at level one).  This \emph{protograph} is recursively appended to trees to obtain the set $\cT_v$.  The algorithm is presented in a somewhat informal way here for clarity; we formalize it slightly in Appendix~\ref{app:complete} in order to prove the following proposition.

\begin{prop}\label{prop:complete}
Algorithm~\ref{alg:badiu} returns the complete reduced set of full binary trees with $v$ internal nodes.
\end{prop}
\begin{IEEEproof}
See Appendix~\ref{app:complete} in the Supplemental Material.
\end{IEEEproof}

\begin{figure}[t]
 \removelatexerror
  \begin{algorithm}[H]
   \caption{Recursive Construction of $\cT_v$}
   \label{alg:badiu}
   {\bf initialize} $\cT_1 = \{\tau\};\,\cT_{k'} = \{\} \,\forall k' = 2,\ldots,v;\, k=2$\;
    \While{$k \leq v$}
    {
        \For{$t \in \cT_{k-1}$}
        {
            Append $\tau$ to the left-most leaf on the lowest level of $t$, and add the new tree to $\cT_k$\;
            Append $\tau$ to the left-most available leaf on the next-to-lowest level of $t$ where possible, and add the new tree to $\cT_k$\;
        }
        $k \leftarrow k+1$\;
    }
  \end{algorithm}
\end{figure}

\begin{figure}[t]
    \centering
%
%
    \includegraphics[angle=0,width=8cm,clip=false]
    {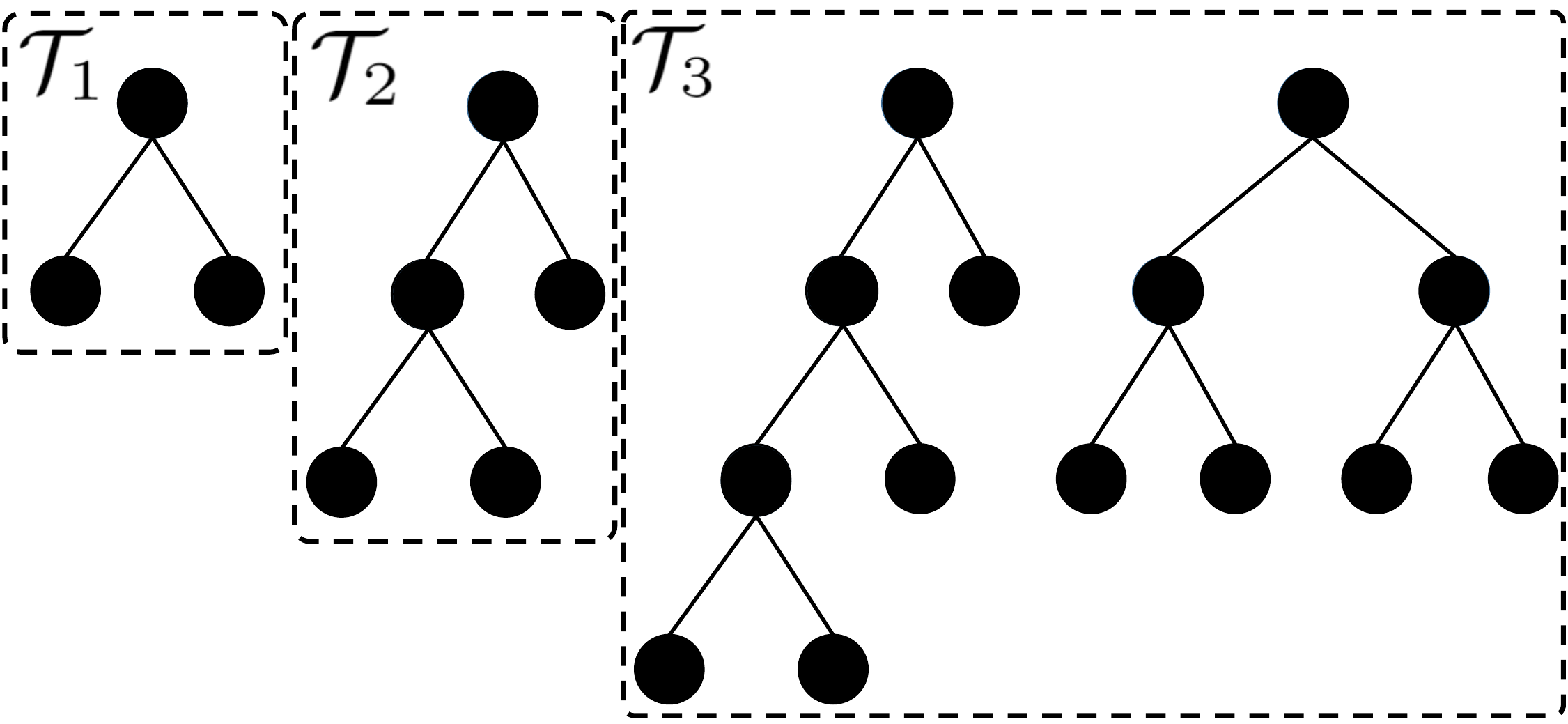}
    \caption{Illustration of the sets $\cT_1$, $\cT_2$, and $\cT_3$ generated by Algorithm~\ref{alg:badiu}.}
    \label{fig:Tv}
\end{figure}

As an example of the output of Algorithm~\ref{alg:badiu}, Fig.~\ref{fig:Tv} shows the sets generated for $v = 1,2,3$.  Note that the tree shown for the set $\cT_1$ is the protograph $\tau$.  The number of protographs contained in a graph of $\cT_v$ is $v$.

\subsection{Enumeration}
As noted above, the number of ordered full binary trees with $v$ internal nodes is given by the Catalan number $c_v$.  The reduced set of $v$-node full binary trees contains significantly fewer elements.  For example, Fig.~\ref{fig:Tv} shows that two trees are contained in $\cT_3$; yet, by considering all orderings of these two trees, we can enumerate five ordered trees ($c_3 = 5$).

Let $T_v$ denote the number of trees in the set $\cT_v$.  From Algorithm~\ref{alg:badiu}, we can infer the relations
\begin{equation}\label{eq:loose_bound}
    T_v \leq 2 T_{v-1} \leq \cdots \leq 2^{v-1} T_1 = 2^{v-1}
\end{equation}
since each step in the for loop \emph{at most} doubles the number of elements in $\cT_k$.  This bound captures the slower exponential growth in the number of trees in the reduced set compared to the set of ordered trees.  Numerical results have shown that the bound overestimates the rate of increase in $v$.  Published results on full binary trees have attempted to obtain generating functions for the number of trees in unordered, unlabelled sets (see, e.g.,~\cite{Murtagh1984} and references therein).  However, it appears that results on the reduced sets that we are interested in remain undiscovered. 

It is possible to obtain a tighter bound on $T_v$ by analyzing Algorithm~\ref{alg:badiu}.  The bound is given as a recurrence relation in the following proposition.

\begin{prop}\label{prop:bound}
The number of trees in $\cT_v$ is upper bounded by
\begin{equation}
    T_v \leq 2 T_{v-1} - \delta_v - \sum_{q=2}^{\floor{\lt(v-1)}} T_{v-2^{{}^q}} 
\end{equation}
where $\delta_v = 1$ if $v$ is a power of two and $\delta_v = 0$ otherwise, and the summation is empty when $v < 5$.
\end{prop}
\begin{IEEEproof}
See Appendix~\ref{app:bound} in the Supplemental Material.
\end{IEEEproof}

\begin{figure}[t]
    \centering
    \includegraphics[width=8cm]{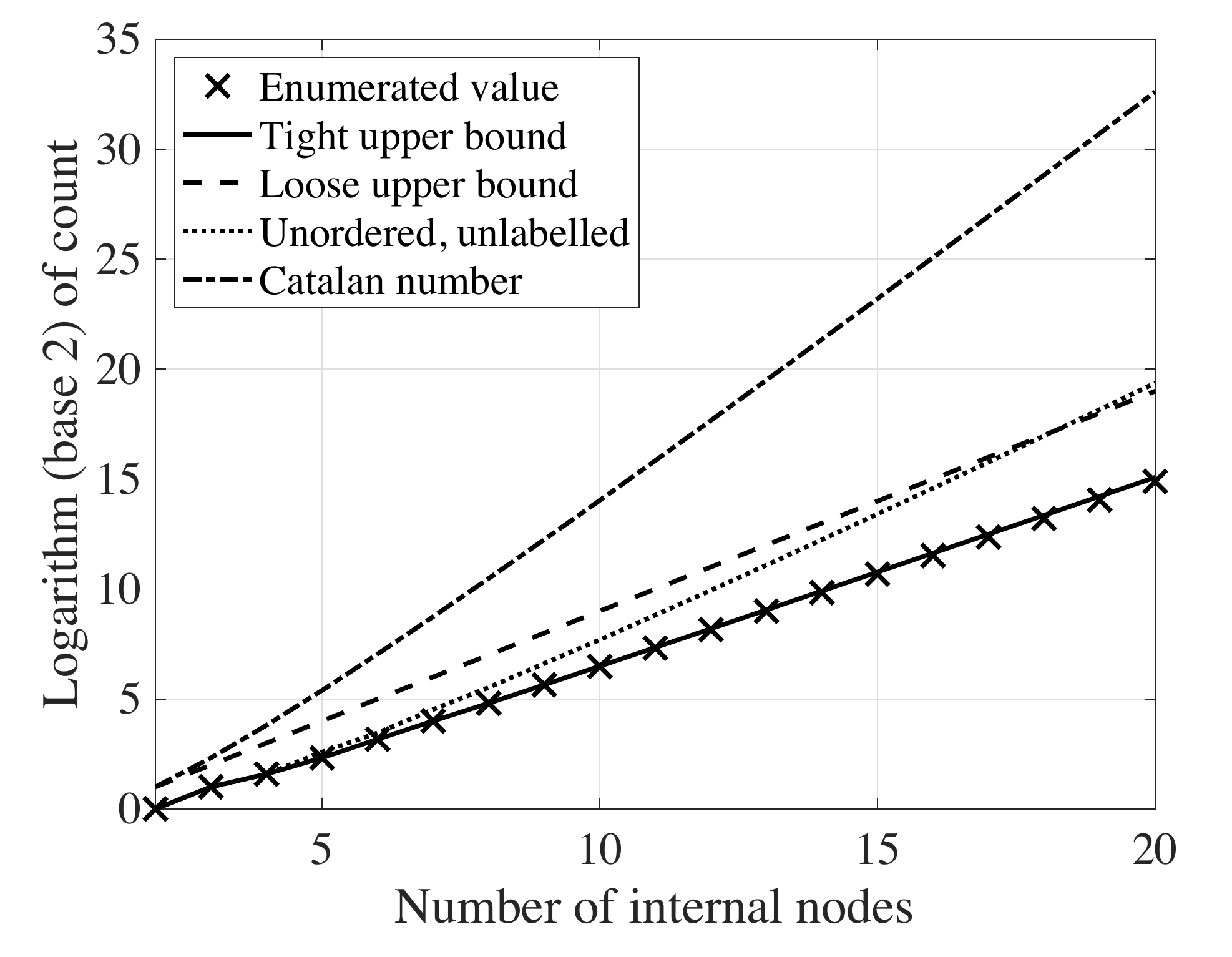}
    \caption{Enumeration of and bounds on the number of full binary trees as a function of the number of internal nodes.  The enumerated result corresponds to Algorithm~\ref{alg:badiu}.  The loose bound is $2^{v-1}$ (cf.~\eqref{eq:loose_bound}).  The tight bound corresponds to the result given in Proposition~\ref{prop:bound}, where the recursion is performed over the bounds on $T_v$ rather than the exact enumerated values. The ``Unordered, unlabelled'' plot corresponds to the enumeration given in~\cite{Murtagh1984}. The number of ordered full binary trees (the Catalan number) is plotted as a reference.}  
    \label{fig:enumeration}
\end{figure}

The accuracy of each of the two bounds given above is illustrated for sets of up to twenty internal nodes in Fig.~\ref{fig:enumeration}.  From the figure, we see that the loose bound slightly overestimates the growth rate of $T_v$.  The recursion is exact up to $v=9$, but slowly diverges for larger $v$, although it clearly remains fairly tight up to $v=20$.  Practically, we will be interested in reasonably small $v$; hence, the recursion is a useful tool for analyzing the IM systems studied in this paper.

\section{Mutual Information Optimization: Relaxation}
\label{sec:rate-relaxed}
We now provide details of new results and methods related to the optimization of the mutual information in OFDM-IM systems.  As noted in Section~\ref{sec:model}, the SAP probabilities are constrained by the binary tree chosen for encoding.  Before we treat these constraints, we will consider the relaxed problem, for which it is assumed that SAPs can be transmitted with any probability.  This will give an upper bound on the achievable rate for the constrained system, and we will use the approaches developed herein to treat that case in Section~\ref{sec:rate-constrained}.

Consider a single set of $N$ subcarriers that adhere to the model described in Section~\ref{sec:model}.  We collect the $N$ received symbols in the vector $\bm{Y} \coloneqq (Y_1,\ldots,Y_N)$.  Furthermore, we collect the $K$ transmitted symbols in the vector $\bm{X} \coloneqq (X_1,\ldots,X_N)$, noting that $X_l$ is nonzero only when subcarrier $l$ is active, as given by the encoded SAP.  Define the SAP probability vector $\mathbf{p}=(p_i)$ and the power vector $\bm{\rho}=(\rho_{li})$. We are interested in the probabilities in $\mathbf{p}$ and transmit powers in $\bm{\rho}$ that maximize the mutual information 
\begin{align}\label{eq:MI_basic}
I(\bm{X};\bm{Y}) &= h(\bm{Y}) - h(\bm{Y}\mid\bm{X}) \nonumber\\
&= h(\bm{Y}) - h(\bm{Z}).
\end{align}

Conditioned on $U=i$, we assume $X_l\sim\mathcal{CN}(0,\rho_{li})$ when $l\in\cS_i$.  Choosing $X_l$ to be Gaussian is not proven to achieve capacity, but the assumption provides a tractable expression.  In this case, the complex random vector $\bm{Y}$ has probability density function (pdf)
\begin{equation}\label{eq:fY}
f_{\bm{Y}}(\mathbf{y}) = \sum_{i\in\mathcal{U}} p_i f_{\bm{Y}\mid U}(\mathbf{y}\mid U=i)
\end{equation}
where
\begin{equation}\label{eq:fYcondX}
f_{\bm{Y}\mid U}(\mathbf{y}\mid U=i) = \prod_{l\in\mathcal{S}_i} f_\text{CN}(y_l;g_l\rho_{li}+\sigma^2) \prod_{m\notin\mathcal{S}_i} f_\text{CN}(y_m;\sigma^2)
\end{equation}
with $f_\text{CN}:\mathbb{C}\to[0,\infty)$ being the complex Gaussian pdf $f_\text{CN}(t;\nu)=e^{-|t|^2/\nu}/(\pi\nu)$ with mean zero and variance $\nu$.

Writing $I(\bm{X};\bm{Y}) = I(\mathbf{p},\bm{\rho},\sigma^2)$, the optimization problem is formulated as
\begin{equation}\label{eq:opt_MI}
\begin{aligned}
    &\underset{\mathbf{p},\bm{\rho}}{\text{maximize}} && I(\mathbf{p},\bm{\rho},\sigma^2) \\
    &\text{subject to} && \sum_{i\in\mathcal{U}} p_i = 1 \\
    &&& \sum_{l\in\mathcal{S}_i}\rho_{li} \leq P,\quad\forall i\in\mathcal{U} \\
    &&& p_i\geq 0,\quad\forall i\in\mathcal{U} \\
    &&& \rho_{li}\geq 0,\quad\forall l\in\mathcal{S}_i,\, i\in\mathcal{U}.
\end{aligned}
\end{equation}
Note that the relaxation alluded to earlier manifests in the simple constraint $\sum_{i\in\mathcal{U}} p_i = 1$.  If we were to consider only probability vectors $\mathbf{p}$ that adhere to the binary-tree encoding methodology, this constraint would be defined differently (see Section~\ref{sec:rate-constrained}).  We now detail several strategies for solving, either approximately or exactly, the optimization problem stated in~\eqref{eq:opt_MI}.

\subsection{Concavity and Numerical Optimization}
The following result that can be used to solve~\eqref{eq:opt_MI} numerically.
\begin{lem}\label{lem:concave}
For fixed $\bm{\rho}$, the problem
\begin{equation}\label{eq:opt_MI_p}
\begin{aligned}
    &\underset{\mathbf{p}}{\text{maximize}} && I(\mathbf{p},\bm{\rho},\sigma^2) \\
    &\text{subject to} && \sum_{i\in\mathcal{U}} p_i = 1 \\
    &&& p_i\geq 0,\quad\forall i\in\mathcal{U}.
\end{aligned}
\end{equation}
is concave.
\end{lem}
\begin{IEEEproof}
See Appendix~\ref{app:concave} in the Supplemental Material.
\end{IEEEproof}
For the special case where $g_l$ is constant for all $l$ and a balanced power distribution is chosen (i.e., $\rho_{li} = \rho$ for all $l$ and $i$), Lemma~\ref{lem:concave} leads to the following.
\begin{prop}\label{prop:awgn}
When the channel gains and transmit powers are constant across frequency, the optimal SAP probability distribution is uniform.
\end{prop}
\begin{IEEEproof}
See Appendix~\ref{app:awgn} in the Supplemental Material.
\end{IEEEproof}

More generally, Lemma~\ref{lem:concave} suggests that it may be economic to solve~\eqref{eq:opt_MI} by employing a block coordinate descent (BCD) approach~\cite{Bertsekas1999}, in which one would alternately maximize the mutual information in either $\mathbf{p}$ or $\bm{\rho}$ while keeping the other vector fixed at the previously obtained optimum value.  The method requires the constraints of the problem to be convex, which is clearly satisfied.  Furthermore, the maximization over each of the vectors $\mathbf{p}$ and $\bm{\rho}$, keeping the other constant, must be unique.  Lemma~\ref{lem:concave} implies this condition is met in part, but it is not clear whether the condition may be violated for the maximization of $I(\mathbf{p},\bm{\rho},\sigma^2)$ over $\bm{\rho}$ for a fixed $\mathbf{p}$ in some parameterizations of $\{g_l\}$ and $\sigma^2$.  Nevertheless, the smoothness of the objective function provides some assurance that a BCD approach will converge to a local extremum.  

One may encounter numerical problems when using the BCD technique to solve~\eqref{eq:opt_MI} since, in general, the evaluation of $I(\mathbf{p},\bm{\rho},\sigma^2)$ requires high-dimensional numerical integration or time-consuming Monte Carlo methods.  In practice, we have found that the BCD method can only be employed to optimize systems with three or four subcarriers per group; larger systems require different approaches.

\subsection{A Lower Bound}
It is possible to obtain an approximate solution to~\eqref{eq:opt_MI} by considering a lower bound on the mutual information rather than the mutual information, itself.  The following proposition provides one such bound.
\begin{prop}\label{prop:jensen}
For transmit powers $\bm{\rho}$ and SAP probabilities $\mathbf{p}$, $I(\mathbf{p},\bm{\rho},\sigma^2)$ satisfies the lower bound  
\begin{equation}\label{eq:jensen}
  I(\mathbf{p},\bm{\rho},\sigma^2) \geq -\ln\left(\sum_{i \in \mathcal{U}}\sum_{j \in \mathcal{U}}  \frac{p_i p_j}{\det(\bm{\Xi}_i+\bm{\Xi}_j)}\right)-N\ln\left( e \sigma^2\right)
\end{equation}
where $\bm{\Xi}_i$ is a diagonal matrix with the $l$th element of the diagonal $\xi_{li}$ satisfying
\begin{equation}
    \xi_{li} = 
    \begin{cases}
    g_l \rho_{li} + \sigma^2,\quad &\text{if } l\in\cS_i, \\
    \sigma^2,&\text{otherwise.} 
    \end{cases}
\end{equation}
\end{prop}
\begin{IEEEproof}
See Appendix~\ref{app:jensen} in the Supplemental Material.
\end{IEEEproof}

The bound given above is a result of Jensen's inequality and is, thus, not particularly tight.  In fact, a slightly different application of the inequality yields a marginally tighter bound~\cite[Th. 2]{Huber2008}.  However, the utility in Proposition~\ref{prop:jensen} is not in the accuracy of the bound, but rather in the ease with which this bound can be optimized over the SAP probabilities.  These optimal probabilities are captured in the following proposition.

\begin{prop}\label{prop:jensen-opt}
Let $\bm{A} = (a_{ij})$ with $a_{ij} = 1/\det(\bm{\Xi}_i+\bm{\Xi}_j)$.  Suppose $\bm{A}$ is nonsingular, and let $\bm{B} = \bm{A}^{-1}$, with $b_{ij}$ denoting the element in the $i$th row and $j$th column of $\bm{B}$.  The SAP probabilities that maximize the lower bound given in~\eqref{eq:jensen} are given by
\begin{equation}\label{eq:pi-jensen}
    p_i = \frac{\left(\sum_{j\in\cU} b_{ij}\right)^+}{\sum_{i\in\cU}\left(\sum_{j\in\cU} b_{ij}\right)^+},\quad \forall i \in \cU
\end{equation}
where $(x)^+ = \max\{x,0\}$.
\end{prop}
\begin{IEEEproof}
See Appendix~\ref{app:jensen-opt} in the Supplemental Material.
\end{IEEEproof}

Note that the probabilities given in~\eqref{eq:pi-jensen} are dependent upon the subcarrier powers.  The BCD approach can be employed in a fairly straightforward manner to compute the power values by alternately computing~\eqref{eq:pi-jensen} for fixed powers, then fixing these probabilities in~\eqref{eq:jensen} and computing the maximizing power values.  Alternatively, one can, in theory, substitute~\eqref{eq:pi-jensen} into~\eqref{eq:jensen} and compute the optimal powers directly.  However, the nonlinear form of~\eqref{eq:pi-jensen} can cause problems using this approach.  

A condition that must be satisfied in order to invoke Proposition~\ref{prop:jensen-opt} is that $\bm{A}$ must be nonsingular.  It is possible that this condition is not met, for example when only a single subcarrier in the set of $K$ active subcarriers is allocated power.  Such cases can typically be dealt with by using other results reported in this section (e.g., the asymptotic results detailed below).  In general, we have found that Proposition~\ref{prop:jensen-opt} is applicable to a wide range of system configurations.

\subsection{Closed-Form Asymptotics}
\label{sec:asymptotic_results}
It is naturally preferable to solve~\eqref{eq:opt_MI} analytically.  To make progress in this direction, we apply the following strategy:
first, we find the probabilities $\mathbf{p}^\star(\bm{\rho})$ that maximize the mutual information for any given values of the transmit powers, i.e., the optimal probabilities are functions of the powers; then, the mutual information $I(\mathbf{p}^\star(\bm{\rho}),\bm{\rho},\sigma^2)$ that corresponds to the optimal probabilities found previously is maximized over the powers in $\bm{\rho}$.  

\subsubsection{Probability Optimization}
To obtain a closed-form expression for the optimal SAP probability distribution as a function of the powers, we first resort to a high-SNR analysis, which gives rise to the following result.

\begin{prop}\label{prop:p_opt_SNRinf}
For fixed powers $\bm{\rho}$, let $I^\star(\bm{\rho},\sigma^2)=\max_{\mathbf{p}} I(\mathbf{p},\bm{\rho},\sigma^2)$ and define the probability values
\begin{equation}\label{eq:prob_optim}
q_i = q_i(\sigma^2) = \frac{\prod_{l\in\mathcal{S}_i}(g_l\rho_{li}+\sigma^2)}{\sum_{j\in\mathcal{U}} \prod_{m\in\mathcal{S}_j}(g_m\rho_{mj}+\sigma^2)}.
\end{equation}
Then, $I(\mathbf{q},\bm{\rho},\sigma^2)$ is a lower bound of $I^\star(\bm{\rho},\sigma^2)$ that is tight in the high SNR regime, i.e., as $\sigma^2\to 0$.
\end{prop}
\begin{IEEEproof}
See Appendix~\ref{app:p_opt_SNRinf} in the Supplemental Material.
\end{IEEEproof}

In addition to a simple, closed-form expression for the optimal SAP probability distribution at high SNR, this result also provides an upper bound on the achievable rate, as stated in the following corollary.

\begin{corollary}\label{cor:UB}
The function $\mu(\sigma^2)\coloneqq\ln \left[\sum_{i\in\mathcal{U}} \prod_{l\in\mathcal{S}_i}\left(\frac{g_l\rho_{li}}{\sigma^2}+1\right) \right]$ is an upper bound on the maximal mutual information $I^\star(\bm{\rho},\sigma^2)$, which is tight as $\sigma^2\to 0$.
\end{corollary}
\begin{IEEEproof}
See Appendix~\ref{app:UB} in the Supplemental Material.
\end{IEEEproof}

We now turn our attention to the low-SNR case, for which we obtain the following beautifully intuitive result, which is a somewhat discrete version of the well known waterfilling principle at low SNR.

\begin{prop}\label{prop:p_opt_SNRlow}
For fixed powers $\bm{\rho}$, let $i^\star=\operatorname{arg\,max}_{i} \sum_{l\in\mathcal{S}_i}\ln\left(\frac{g_l\rho_{li}}{\sigma^2}+1\right)$, i.e., $i^\star$ corresponds to the group of $K$ strongest subcarriers. Then, the index probabilities
\begin{equation}\label{eq:prob_optim_lowSNR}
r_i = 
\begin{cases}
1,\quad &\text{if } i=i^\star, \\
0,&\text{otherwise}, 
\end{cases}
\end{equation}
maximize the mutual information at low SNR, which satisfies the asymptotic equivalence $I^\star(\bm{\rho},\sigma^2)\sim \sum_{l\in\mathcal{S}_{i^\star}}\ln\left(\frac{a_{li^\star}}{\sigma^2}+1\right)$, as $\sigma^2\to\infty$.
\end{prop}
\begin{IEEEproof}
See Appendix~\ref{app:p_opt_SNRlow} in the Supplemental Material.
\end{IEEEproof}

\subsubsection{Power Optimization}
Propositions~\ref{prop:p_opt_SNRinf} and~\ref{prop:p_opt_SNRlow}  and Corollary~\ref{cor:UB} yield closed-form expressions for the mutual information, which depend upon the powers $\bm{\rho}$.  As a result, these expressions can be used to develop optimal power allocation rules in the high and low-SNR regimes.  It turns out that the optimal rules follow our conventional understanding of power allocation in OFDM systems, as formalized in the following proposition.

\begin{prop}\label{prop:rho_opt_SNRinf_SNRlow}
For high SNR (as $\sigma^2\to 0$), allocating powers for the subcarriers of each SAP according to the waterfilling strategy is optimal under power constraints for each pattern.  For low SNR (as $\sigma^2\to \infty$), allocating powers according to the waterfilling strategy is optimal.
\end{prop}
\begin{IEEEproof}
See Appendix~\ref{app:rho_opt_SNRinf_SNRlow} in the Supplemental Material.
\end{IEEEproof}

The waterfilling result for the low-SNR case is somewhat unsurprising given that Proposition~\ref{prop:p_opt_SNRlow} indicates the mutual information expression is the same as that for OFDM with only $K$ active subcarriers.  On the other hand, the optimality of waterfilling at high SNR is not immediately obvious from Corollary~\ref{cor:UB}.  These results lead us to a simple mutual information optimization strategy for $\mathbf{p}$ and $\bm{\rho}$ at high and low SNR: one should perform waterfilling power allocation for each subcarrier pattern and then compute the corresponding probabilities according to Proposition~\ref{prop:p_opt_SNRinf} or~\ref{prop:p_opt_SNRlow} and select the result that maximizes the objective.

\section{Achievable Rate Optimization: Constrained}
\label{sec:rate-constrained}
We now consider a more practical rate optimization problem that is effectively the same as~\eqref{eq:opt_MI} but with a nonlinear constraint on the probabilities $\{p_i\}$.  As discussed in Sections~\ref{sec:model} and~\ref{sec:trees}, SAP probabilities depend on two things: (1) the full binary tree that corresponds to the bit-to-SAP mapping operation, and (2) the ordering of the SAP-to-leaf assignment.  

Let $\cP_v$ denote the set of feasible probability vectors of length $C$ that can be constructed by considering all non-redundant SAP-to-leaf assignments for all binary trees in $\cT_v$ \emph{including null assignments}.  For example, for $C=4$, $\cP_2$ is constructed by considering all mappings of three (out of four) SAPs to two leaves on the second level and one leaf on the first level of the single tree in $\cT_2$ (cf. Fig.~\ref{fig:Tv}).  There are $3!/2! = 3$ mappings of three SAPs to the leaves, and $\binom{4}{3} = 4$ ways of choosing the \emph{active} SAPs.  The inclusion of null assignments in this way ensures we consider the case of not using some SAPs that may correspond to poor channel conditions.  Under this definition of $\cP_v$, the number of elements (probability vectors) in $\cP_v$ is
\begin{equation}
    P_v = \binom{C}{v+1}\sum_{t\in\cT_v}\frac{(v+1)!}{n_{t1}!\cdots n_{tv}!}
\end{equation}
where $n_{tq}$ denotes the number of leaves at level $q$ in tree $t$.  We further define the union $\cP = \cup_{v=0}^{C-1}\cP_v$, where $\cP_0$ consists of the $C$ vectors with one element equal to one and the rest equal to zero.

The constrained optimization problem can now be formulated as
\begin{equation}\label{eq:opt_MI_constrained}
\begin{aligned}
    &\underset{\mathbf{p},\bm{\rho}}{\text{maximize}} && I(\mathbf{p},\bm{\rho},\sigma^2) \\
    &\text{subject to} && \sum_{i\in\mathcal{U}} p_i = 1 \\
    &&& \mathbf{p}\in \cP \\
    &&& \sum_{l\in\mathcal{S}_i}\rho_{li} \leq P, \quad\forall i\in\cU \\
    &&& \rho_{li}\geq 0,\quad \forall l\in\mathcal{S}_i,\, i\in\mathcal{U}.
\end{aligned}
\end{equation}
We propose two methods of solving this problem here: an enumerative approach, and a projection from the relaxation.

\subsection{Enumerative Approach}
This approach is based on the enumeration of all possible probability distributions of the SAPs, i.e., all $\bp\in\cP$. The allocated powers $\bm{\rho}$ are optimized for each probability distribution $\bp$.  The pair $(\bp,\bm{\rho})$ that yields the highest mutual information is the solution to the problem stated in~\eqref{eq:opt_MI_constrained}.  

For a given probability distribution $\mathbf{p}$, the power allocation problem may not be solved analytically.  In this case, one can invoke the asymptotic results stated in Proposition~\ref{prop:rho_opt_SNRinf_SNRlow} to obtain the power values.  First, the waterfilling power allocation solution would be calculated for each SAP.  Then, the distribution $\bp\in\cP$ that maximizes the mutual information would be chosen.

\subsection{Projection from the Relaxation}
A much more computationally efficient method of treating~\eqref{eq:opt_MI_constrained} can be developed by first considering the relaxation studied in the previous section. First, we relax the constraint $\bp\in\cP$ to find a solution to~\eqref{eq:opt_MI}.  Any of the approaches used in Section~\ref{sec:rate-relaxed} can be applied.  We let $\bp'$ denote the probability distribution computed in this step.  We then project $\bp'$ onto the feasible vector $\bp^\star\in\cP$ and take this to be the partial solution to~\eqref{eq:opt_MI_constrained}.  The power allocation vector $\bm{\rho}^\star$ is then computed to maximize the mutual information.

The projection of the relaxed solution $\mathbf{p}'$ onto a point in the set $\cP$ can be accomplished efficiently by using the Huffman coding algorithm~\cite{Huffman1952}.  To this end, we interpret the elements of $\bp'$ as source symbol probabilities, then generate a full binary tree according to the Huffman algorithm.  As discussed in Section~\ref{sec:model}, SAPs associated with a leaf node in the tree at level $q$ will be transmitted with probability $1/2^q$.  Hence, the probabilities in $\bp'$ are replaced with the corresponding probabilities derived from the tree structure to yield a candidate for $\bp^\star$.  

It is important to note that this basic approach will only yield trees (and associated probability distributions) with $C$ leaves, i.e., the algorithm maps $\bp'$ to $\cP_C$ only.  To ensure we consider mappings to all points in $\cP$, we require a slightly modified approach.  The full details of the complete projection algorithm are given in Algorithm~\ref{alg:liu}, and an example depicting how the algorithm works is shown in Fig.~\ref{fig:proj}.  The function $\operatorname{sort}(\cdot)$ in Algorithm~\ref{alg:liu} arranges the set of $C$ arguments in decreasing order; $\operatorname{unsort}(\cdot)$ performs the inverse mapping (again, acting on $C$ elements).  The function $\operatorname{Huffman}(\cdot)$ takes a set of ``source probabilities'' and returns the corresponding set of depths, or path lengths from the root to a given leaf.  Finally, the function $\operatorname{dist}(\mathbf{p}_1,\mathbf{p}_2)$ computes the distance between the discrete probability distributions $\mathbf{p}_1$ and $\mathbf{p}_2$.  In the next section, we consider three distance measures: Euclidean distance, for which
\begin{equation}
    \operatorname{dist}(\mathbf{p}_1,\mathbf{p}_2) \coloneqq \Vert \mathbf{p}_1 - \mathbf{p}_2 \Vert^2
\end{equation}
Kullback–Leibler (KL) divergence, for which
\begin{equation}
    \operatorname{dist}(\mathbf{p}_1,\mathbf{p}_2) \coloneqq \sum_{i\in\cU} p_{1i} \ln \frac{p_{1i}}{p_{2i}}
\end{equation}
and total variation distance, for which
\begin{equation}
    \operatorname{dist}(\mathbf{p}_1,\mathbf{p}_2) \coloneqq \max_{i\in\cU}\{\vert p_{1i} - p_{2i} \vert\}.
\end{equation}

Algorithm~\ref{alg:liu} is heuristic.  It is not guaranteed to produce the solution to~\eqref{eq:opt_MI_constrained}.  However, results have shown it performs very well in practical scenarios (see Section~\ref{sec:numerics}).

\begin{figure}[t]
 \removelatexerror
  \begin{algorithm}[H]
   \caption{Projection from the Relaxation}
   \label{alg:liu}
   {\bf initialize} Compute $\bp'$ by (approximately) solving~\eqref{eq:opt_MI}, set $\bar{\bp} = \operatorname{sort}(\bp')$, set $k = 1$\;
    \While{$k \leq C$}
    {
        $s_k = \sum_{l=1}^{C-k+1} \bar{p}_{l}$\;
        $\{q_1,\ldots,q_{C-k+1}\} = \operatorname{Huffman}(\bar{p}_{1}/s_k,\ldots,\bar{p}_{C-k+1}/s_k)$\;
        $\bp_k^\star = \operatorname{unsort}(1/2^{q_1},\ldots,1/2^{q_{C-k+1}},0,\ldots,0)$\;
        $k \leftarrow k+1$\;
    }
    {\bf return} $\bp^\star = \arg\min_{\mathbf{t}\in\{\bp_1^\star,\ldots,\bp_C^\star\}} \operatorname{dist}(\mathbf{t},\bp')$\;
  \end{algorithm}
\end{figure}

\begin{figure}
\centering
\includegraphics[width=8cm,clip=false]
{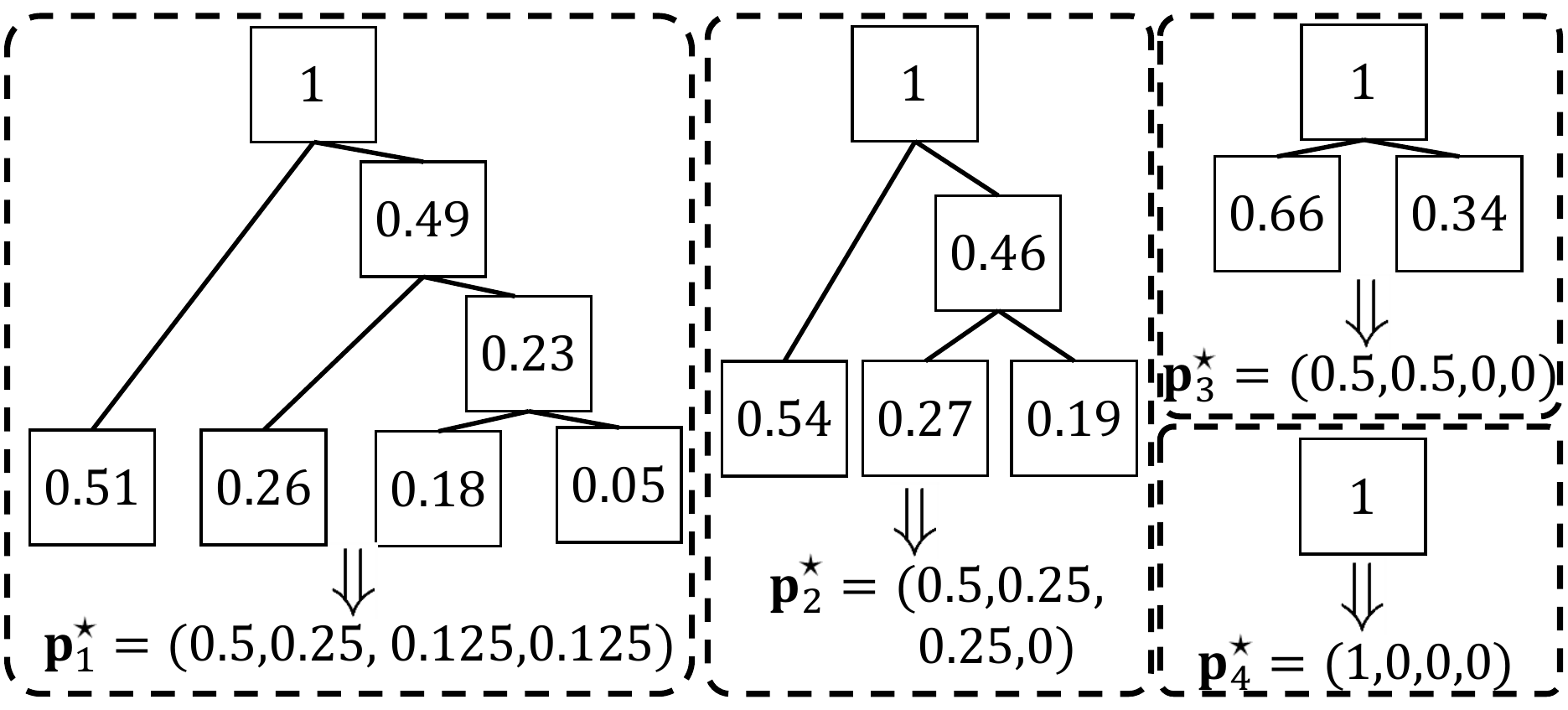}
\caption{Illustration of Algorithm~\ref{alg:liu} for a system with $N=4$ and $K=1$, where $\bp'=(0.51,0.26,0.18,0.05)$. The algorithm constructs four full binary trees according to the input distributions $(0.51,0.26,0.18,0.05)$, $(0.54,0.27,0.19)$, $(0.66,0.34)$, and $(1)$, respectively. The output probability distributions of the algorithm are as follows: $(0.5,0.25,0.25,0)$ for the Euclidean distance metric, $(0.5,0.25,0.125,0.125)$ for the KL divergence metric, and $(0.5,0.25,0.25,0)$ for the total variation distance metric.}
\label{fig:proj}
\end{figure}

\section{Numerical Results}\label{sec:numerics}
In this section, we present a numerical analysis of the methods described above.  We begin with a discussion of the mutual information.  We then give a brief analysis of the error rate of the systems described herein.  In what follows, we define $SNR \coloneqq P/(N\sigma^2)$, which can be interpreted as the average \emph{transmit} SNR per subcarrier.  All mutual information results are given in units of nats, and all curves were obtained via Monte Carlo sampling when closed-form expressions were not available.  For all systems, we set $N=4$ and $K=2$.  It should be noted, however, that we also performed extensive simulations for $(N,K)=(6,4)$ and $(N,K)=(8,6)$, and observed very similar trends to the case where $N=4$ and $K=2$.  We have included some of these results in the Supplemental Material (see Appendix~\ref{app:results}).

\subsection{Mutual Information}
\begin{figure}[t]
    \centering
    \includegraphics[width=8cm]{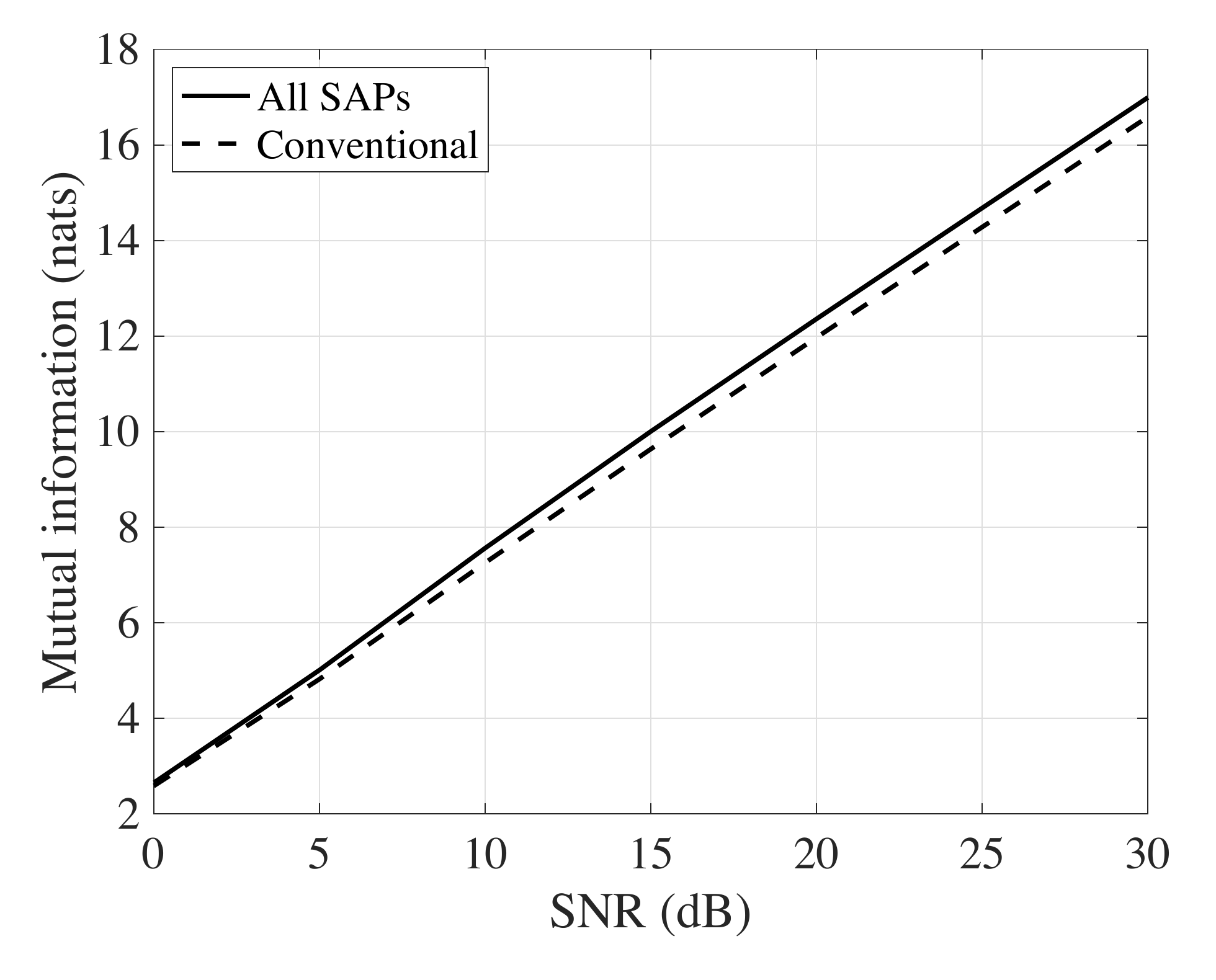}
    \caption{Mutual information vs. SNR for equal channel gains and uniform power allocation.}  
    \label{fig:equal-channel-gains-relaxed}
\end{figure}

We begin with a simple case.  Consider a system operating in AWGN with equal channel gains (i.e., $g_l = 1$ for all~$l$).  Proposition~\ref{prop:awgn} states that the optimal SAP probability distribution in this system is uniform.  Adopting this result, Fig.~\ref{fig:equal-channel-gains-relaxed} shows the mutual information for an OFDM-IM system that uses all six SAPs (each with probability $1/6$) compared to one that limits the number of utilized SAPs to four where uniform power allocation is applied.  The small improvement offered by the former approach simply arises as a result of the additional SAPs that are used\footnote{One would expect an improvement when finite constellations are used for signalling, rather than Gaussian signals.  However, this study is beyond the scope of this paper.}.  However, we note that a uniform SAP distribution is infeasible given a uniform binary source\footnote{This fact has typically been ignored in the literature to date.}.

\begin{figure*}[!t]
    \centering
    \begin{subfigure}[t]{0.5\textwidth}
        \centering
        \includegraphics[width=8cm]{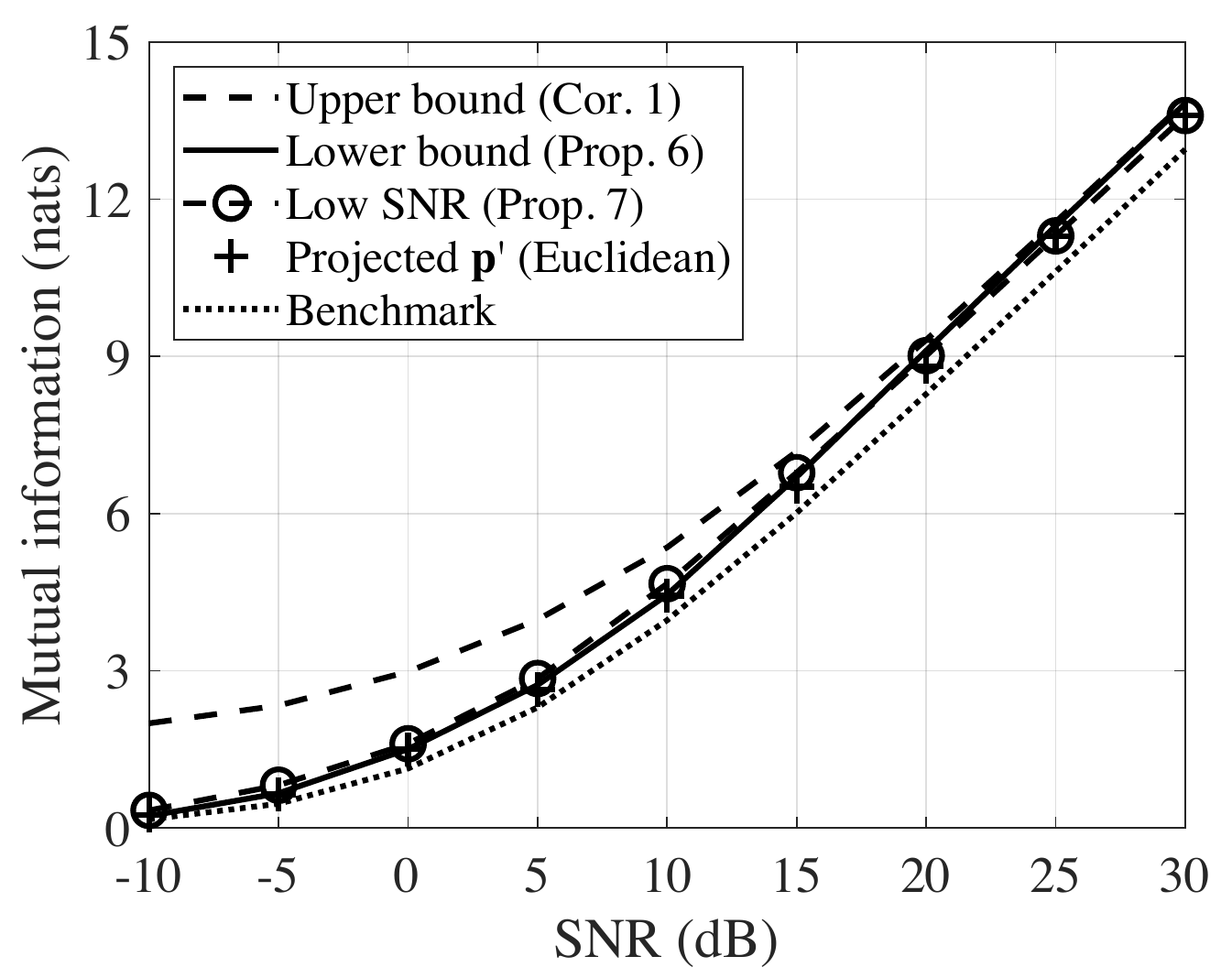}
        \caption{$\eta=0.2$}
        \label{fig:etap2}
    \end{subfigure}%
~
    \begin{subfigure}[t]{0.5\textwidth}
        \centering
        \includegraphics[width=8cm]{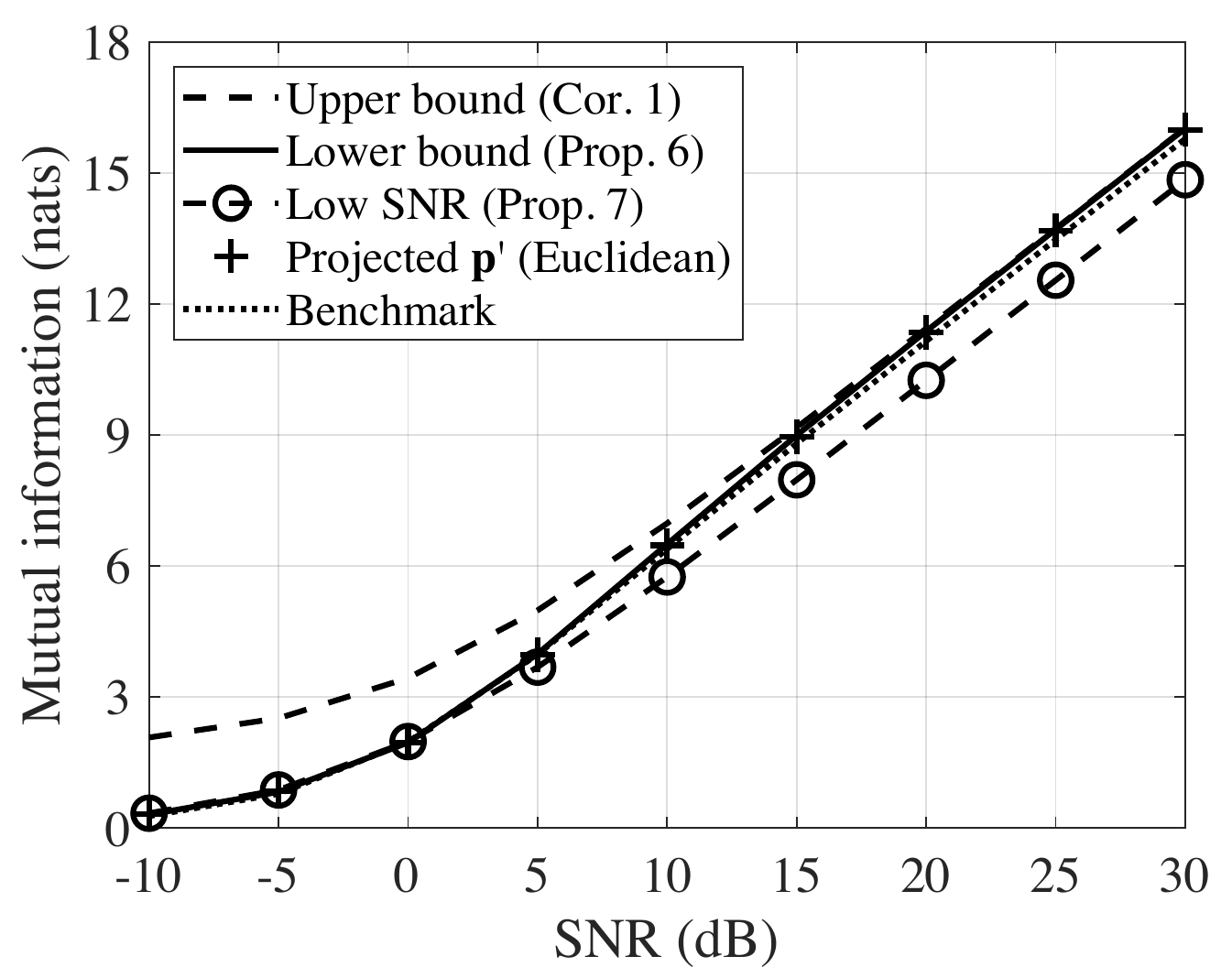}
        \caption{$\eta=0.7$}
        \label{fig:etap7}
    \end{subfigure}
    \caption{Mutual information vs. SNR for $\eta=0.2$ and $\eta=0.7$, where $g_l=\eta^{l-1}$ for $l = 1,\ldots,N$.  Waterfilling power allocation is used for all results except for the benchmark, which uses uniform power allocation.  The curve labelled ``Projected $\bp'$ (Euclidean)'' gives an achievable rate when all SAPs are allowed to be utilized and power allocation is employed.}
    \label{fig:eta}
\end{figure*}

To better understand the advantages that can be brought by utilizing all SAPs along with the binary-tree encoding strategy, we now analyze the case where the channel gains are defined by $g_l=\eta^{l-1}$, $\forall~l\in\{1,\ldots,N\}$ for some $\eta \in (0,1)$.  We assume full channel knowledge is available at the transmitter, so that SAP probabilities and power allocation can be optimized.  Fig.~\ref{fig:etap2} shows the mutual information for $\eta = 0.2$, and Fig.~\ref{fig:etap7} gives results for $\eta = 0.7$.  In both figures, the different curves represent different SAP probability assignment strategies and bounds.  The first three curves exhibit the mutual information computed by using the respective analytic results.  The fourth curve (``Projected $\bp'$ (Euclidean)'') illustrates the mutual information attained by employing Algorithm~\ref{alg:liu}.\footnote{Curves corresponding to other distance functions are not shown because they yield results that are nearly identical to the Euclidean case.}  In this case, $\bp'$ is computed by using the analytic form given in Proposition~\ref{prop:p_opt_SNRinf}.  Note that this curve represents an achievable rate for OFDM-IM systems that utilize all SAPs.  For all curves other than the ``Benchmark'', waterfilling power allocation is employed, since this approach is optimal at high and low SNR in the relaxed setting (cf.~Proposition~\ref{prop:rho_opt_SNRinf_SNRlow}).  The benchmark curve relates to a standard OFDM-IM system where the four SAPs, chosen according to the lexicographic principle discussed in~\cite{Dang2018}, are transmitted with equal probability and uniform power allocation\footnote{Note that the curves that employ waterfilling power allocation and SAP probability optimization include the case where only four SAPs may be chosen, which would correspond to the benchmark curve but where waterfilling is employed.  Such a selection, if deemed to be optimal, would naturally arise through the SAP probability optimization procedure.  As a result, the true benchmark only utilizes uniform power allocation here.}. 

In Fig.~\ref{fig:etap2}, we see that the (relaxed) lower bound of Proposition~\ref{prop:p_opt_SNRinf} and the low-SNR result of Proposition~\ref{prop:p_opt_SNRlow} are similar, and that convergence to the upper bound of Corollary~\ref{cor:UB} occurs at high SNR.  Moreover, the fully constrained result (where $\bp \in\cP$) denoted by the ``$\times$'' markers is very close to the analytic curves corresponding to the relaxed optimization.  The benchmark curve is noticeably lower than all results that offer SAP probability optimization and power allocation.  This was also seen in simulations for $(N,K)=(6,4)$ and $(N,K)=(8,6)$ systems (not shown). Turning our attention to Fig.~\ref{fig:etap7}, we see that the advantages offered by optimization diminish for less variable channel conditions.  The optimized scenario (``Projected $\bp'$ (Euclidean)'') still offers an advantage that saturates the upper bound from mid-to-high SNR values, but it is marginal.  

\begin{figure}[t]
    \centering
    \includegraphics[width=8cm]{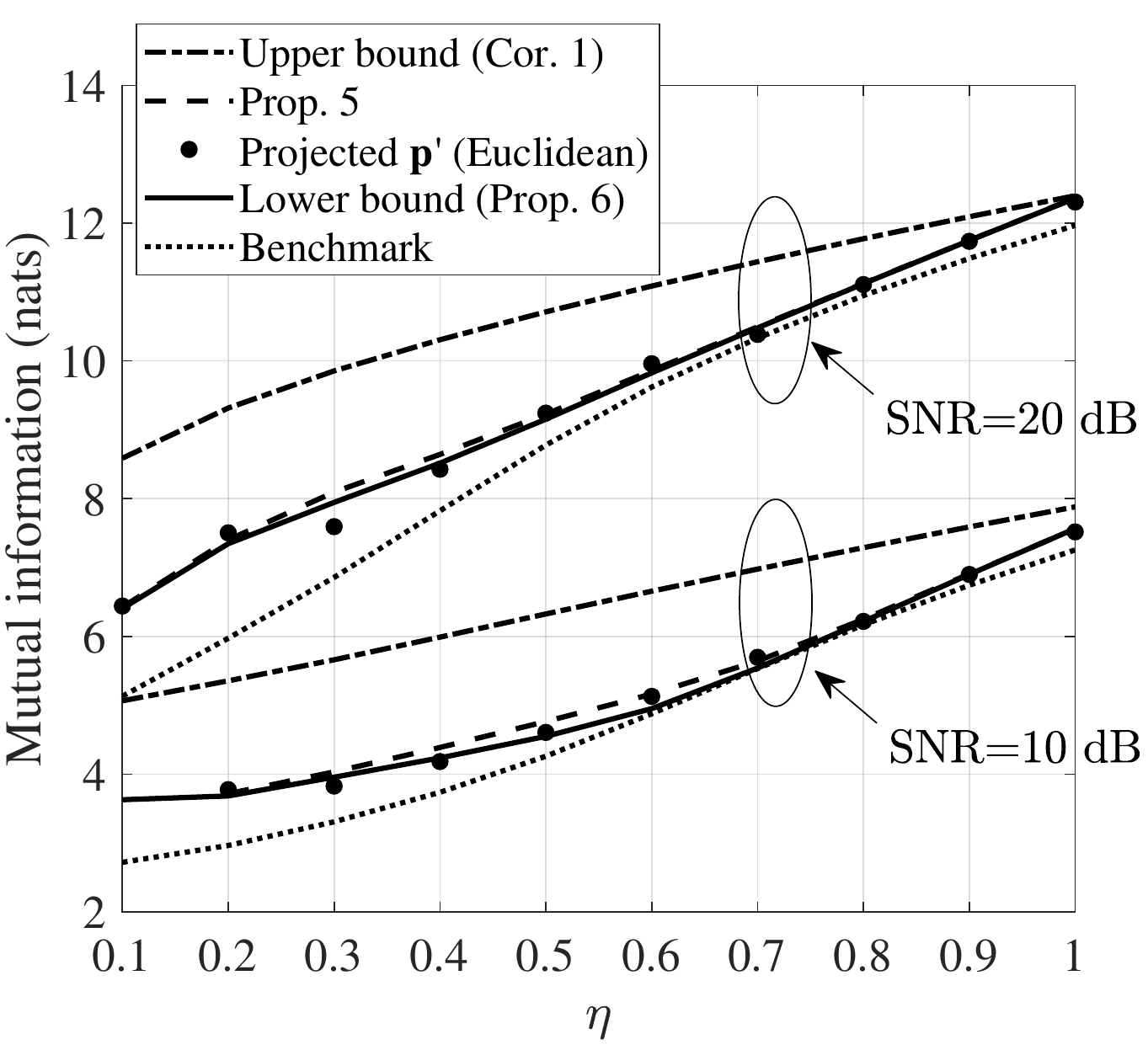}
    \caption{Mutual information vs. $\eta$, where $g_l=\eta^{l-1}$ for $l = 1,\ldots,N$.  The curve labelled ``Projected $\bp'$ (Euclidean)'' gives an achievable rate when all SAPs are allowed to be utilized and power allocation is employed.}  
    \label{fig:var-eta}
\end{figure}

For this simple system ($N=4$ and $K=2$), the results shown in Fig.~\ref{fig:eta} point to a need to understand how frequency selectivity affects performance.  To this end, Fig.~\ref{fig:var-eta} illustrates the mutual information as a function of $\eta$.  The first, fourth, and fifth curves relate to those with the same labels shown in Fig.~\ref{fig:eta}.  The second curve shows the mutual information attained by using the probabilities given in Proposition~\ref{prop:jensen-opt}.  The third curve (``Projected $\bp'$ (Euclidean)'') illustrates the mutual information attained by employing Algorithm~\ref{alg:liu}, where $\bp'$ is computed using Proposition~\ref{prop:jensen-opt}.  Again, waterfilling is used for all systems except the benchmark, where uniform power allocation is used.  The advantages offered by optimization in highly frequency selective channels are apparent in this example\footnote{Again, similar results were observed for $(N,K)=(6,4)$ and $(N,K)=(8,6)$ systems.}.  It is observed that one data point is missing for the $SNR = 10~\text{dB}$ curves related to Proposition~\ref{prop:jensen-opt}.  This omission results from the fact that the matrix $\bm{A}$ in Proposition~\ref{prop:jensen-opt} is singular for the corresponding paramterization.  Hence, for this point, one would choose a different method of obtaining $\mathbf{p}'$ in the initialization step of Algorithm~\ref{alg:liu}.  To conclude this discussion, it is important to note that Algorithm~\ref{alg:liu} roughly achieves the same mutual information promised by the analytic lower bound.  The upper bound is only tight at high SNR; hence, it is not particularly tight for most $\eta$ values in this figure.

\subsection{Block Error Rate}

\begin{figure}[!t]
    \centering
    \begin{subfigure}[t]{0.5\textwidth}
        \centering
        \includegraphics[width=8cm]{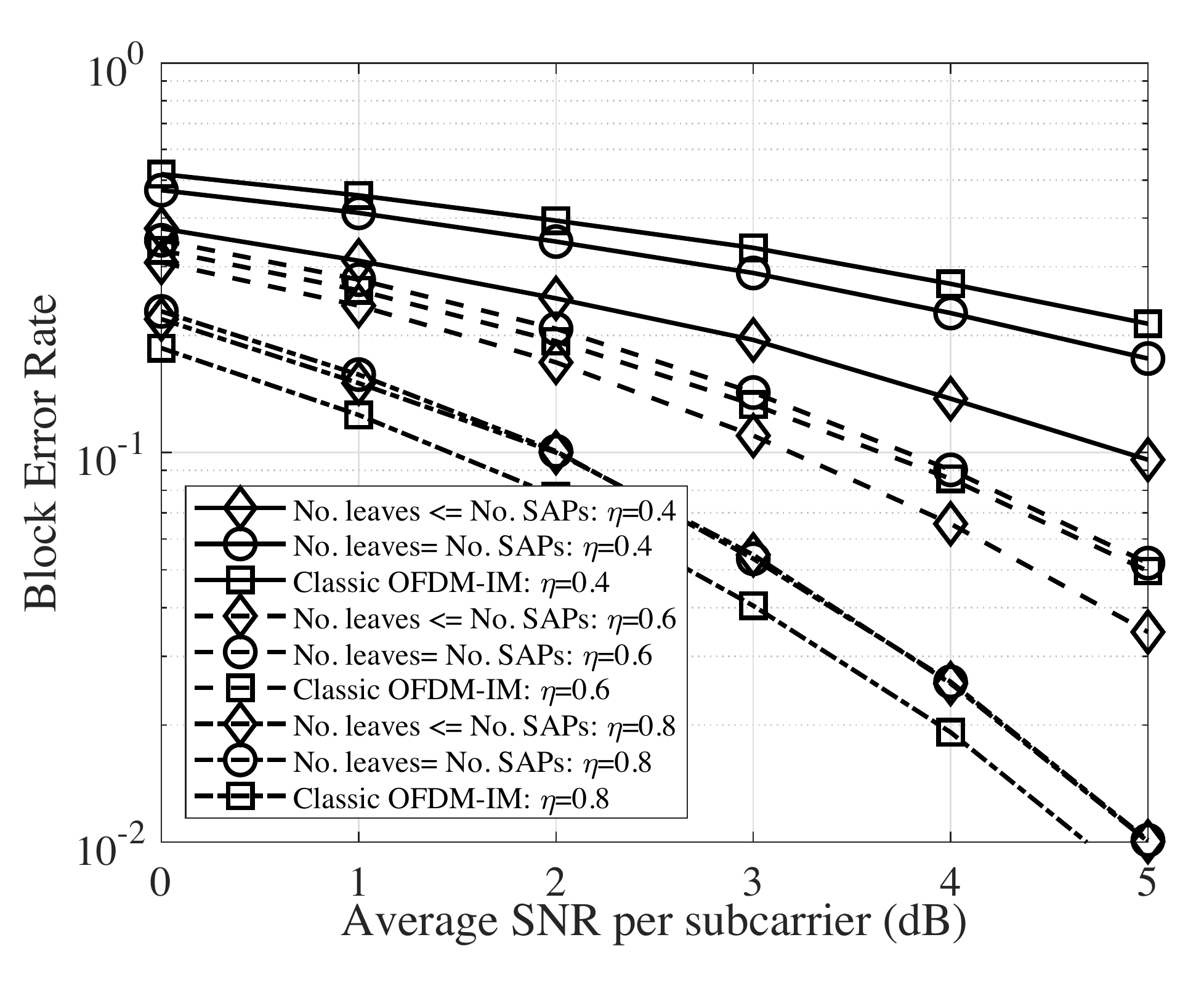}
        \caption{BPSK}
    \end{subfigure}%
\\
    \begin{subfigure}[t]{0.5\textwidth}
        \centering
        \includegraphics[width=8cm]{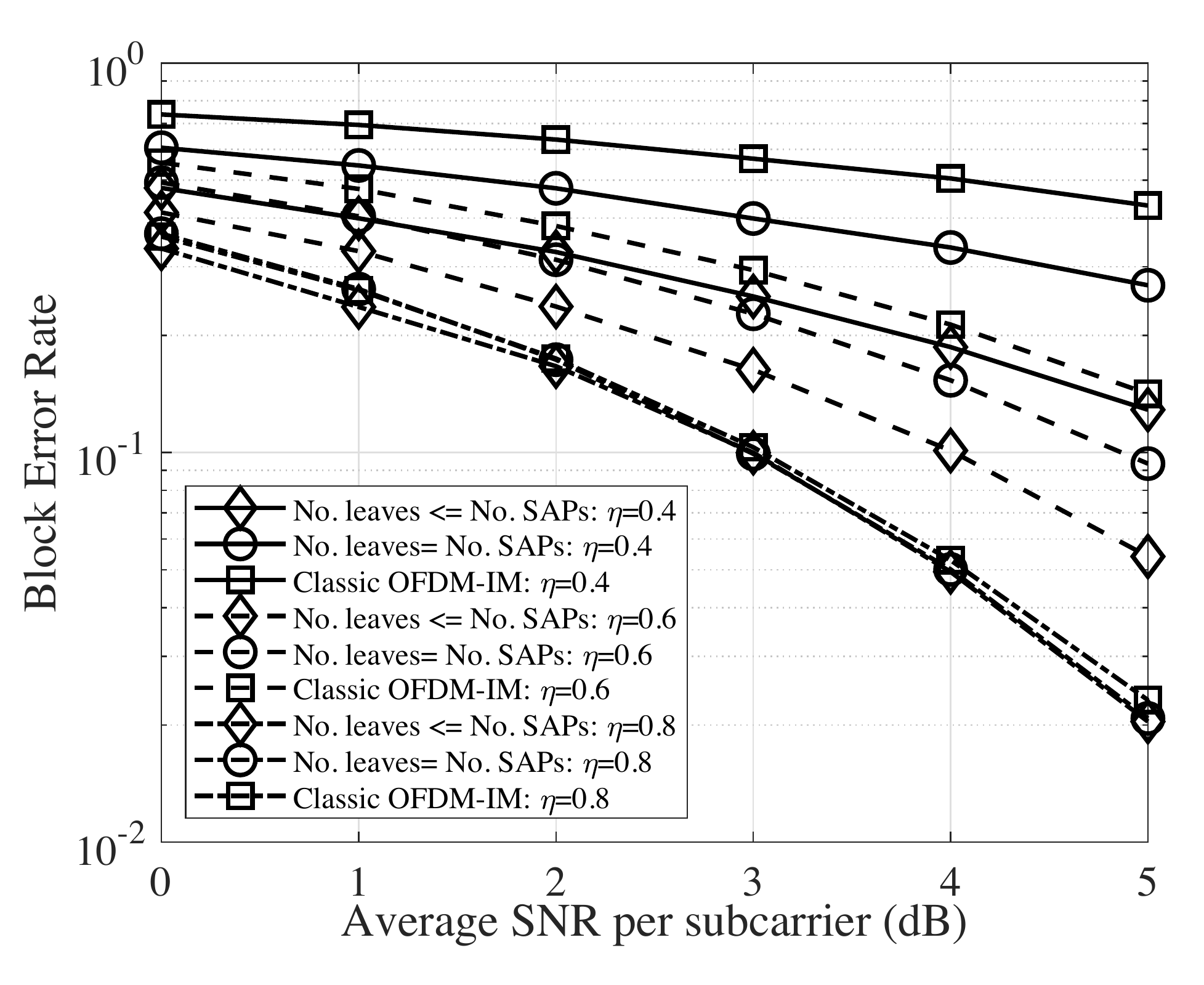}
        \caption{QPSK}
    \end{subfigure}
    \caption{BLER vs. average SNR per subcarrier.}
    \label{BLER}
\end{figure}

Apart from achievable rate, error performance is another key performance metric of a communication system. It should be apparent that a scheme that is designed to maximize the achievable rate does not necessarily optimize the error performance. Nevertheless, it is important to consider the effects that the designs detailed in Sections~\ref{sec:rate-relaxed} and~\ref{sec:rate-constrained} have on the error rate. Note that, because the lengths of the bit sequences encoded as SAPs and modulated signals are variable, the measurement of bit errors would be difficult to assess in a standardized manner. Therefore, to maintain brevity and clarity, we choose to evaluate the block-error rate (BLER) instead of bit-error rate for the~\cite{8241721,8358694,8476574}. Here, a block is a group of $N$ subcarriers.

For simplicity and optimality, we adopt the maximum likelihood (ML) detection scheme at the receiver to estimate the received signal vector $\bm{Y}$ consisting of received modulated symbols and nulls on $N$ independent subcarriers\footnote{Independence can be assumed by considering the case where a subcarrier-level block interleaver is employed~\cite{Xiao2014}.}.  We assume that channel knowledge is available at both transmitter and the receiver.  The estimated signal vector $\hat{\bm{X}}$ satisfies
\begin{equation}
    \hat{\bm{X}}=\underset{\dot{\bm{X}}}{\arg\min}\left\lbrace \bm{Y}-\bm{G}\dot{\bm{X}}\right\rbrace
\end{equation}
where $\dot{\bm{X}}$ is a candidate transmit vector (obtained by following the model described in Section~\ref{sec:model}) and $\bm{G}=\mathrm{diag}\{\sqrt{g_1}e^{j\theta_1},\dots,\sqrt{g_N}e^{j\theta_N}\}$ is the diagonal channel coefficient matrix. We write the BLER conditioned on the transmitted signal vector $\bm{X}$ as $\mathsf{P}_{\mathrm{BLER}}(\bm{X}\vert \bm{G})=\mathbb{P}(\bm{X}\neq\hat{\bm{X}}\vert \bm{G})$. Averaging over $\bm{X}$ gives our measure of interest: $\bar{\mathsf{P}}_{\mathrm{BLER}}(\bm{G})=\mathbb{E}[{\mathsf{P}}_{\mathrm{BLER}}\left(\bm{X}\vert \bm{G}\right)]$, which is now only dependent on the channel state captured in $\bm{G}$.  This measure is useful for evaluating performance in a slow-fading environment.  It also allows us to observe how channel variations affect error performance.

Following the simulation setting for the mutual information analysis, we configured the channel gains to be $g_l=\eta^{l-1}$ and let $\{\theta_l\}$ be uniformly distributed over $[0,2\pi)$, $\forall~l\in\{1,\dots,N\}$. We also normalized $\sigma^2=1$ for the noise power and let $N=4$ and $K=2$ as an example. We do not apply rate adaptation in this study; consequently, we employ a uniform power allocation scheme in all simulations related to error performance. This allows us to focus on the effect that binary-tree optimization (i.e., bit-to-SAP optimization) has on performance.  We numerically examined the BLER for OFDM-IM with the SAP probability distribution optimized under two conditions. The first condition only requires the number of leaves in the binary tree that defines bit-to-SAP mapping to be equal to or smaller than the number of SAPs. The second condition restricts the encoder to only consider full binary trees with $C$ leaves, which reduces the achievable rate at low SNR.  Also, the classic OFDM-IM scheme studied in~\cite{Basar2013} was adopted as a benchmark. We adopt the lexicographic codebook design for the classic OFDM-IM system to select four out of six SAPs for comparison purposes~\cite{Dang2018}.  The numerical results are presented in Fig.~\ref{BLER}, which were obtained by collecting $10^3$ block-error events for each SNR point (subject to random additive white Gaussian noise).

In Fig.~\ref{BLER}, it is apparent that the rate-optimized OFDM-IM systems designed according to the first condition outperform those designed under the second condition. This behavior correlates with the fact that the first condition is less restrictive than the second.  Perhaps more interestingly, we note that the rate-optimized system does not always outperform the classical OFDM-IM scheme.  When signals are subject to deep fading, the rate-optimized system designed according to the first condition may only utilize a single SAP consisting of the best $K$ subcarriers, and the system is reduced to OFDM with $K$ active subcarriers. This system is capable of performing better than those that encode information in the index domain as well as signal space, since block errors arising from incorrect SAP decoding do not occur.

\section{Conclusions}\label{sec:conclusions}
In this paper, we provided a thorough treatment of the rate-optimization problem for OFDM-IM systems with channel knowledge at the transmitter.  We cast the problem as one of mapping bit sequences to activation patterns, which enabled us to utilize a binary tree formalism for algorithm development and analysis.  To this end, we presented new results on full binary trees, both in terms of algorithmic construction and enumeration.  We also reported a number of new analytic bounds and asymptotic results related to the relaxed mutual information optimization problem where SAP probabilities can take any values in the interval $[0,1]$ subject to a sum probability constraint.  We then used the results pertaining to the relaxed problem to develop a heuristic algorithm for obtaining a feasible solution to the constrained problem.  Numerical results indicate that this solution is nearly optimum (relative to the relaxed upper bound), particularly in the low and high SNR regimes, and the optimized approach is capable of offering a rate advantage over the conventional OFDM-IM benchmark of~\cite{Basar2013}.

A number of open problems remain.  First, it is not clear whether an analytic form for the optimal power values exists for all SNR values; only low and high SNR results were reported here.  In fact, it is not readily apparent that the mutual information is concave in the power vector $\bm{\rho}$; hence, a general analytic form may not be forthcoming.  As an alternative, it would be preferable to develop an efficient numerical approach to solving the relaxed optimization problem.  The use of BCD was briefly discussed here, but further work is needed to determine whether this method would be a viable solution in practice.  It is also not known whether the heuristic projection algorithm (Algorithm~\ref{alg:liu}) is, in fact, optimal in some sense.  Finally, and more generally, only rate-optimization for uniform sources was considered; it would be fruitful to study the BLER-minimization problem as well as nonuniform sources and non-Gaussian signalling (i.e., finite signal constellations).

\section*{Acknowledgment}
The authors wish to thank Dr. Hachem Yassine and Mr. Jinchuan Tang for their helpful suggestions regarding the proof of Proposition~\ref{prop:awgn} and the structuring of numerical simulations.

\bibliographystyle{IEEEtran}
\bibliography{bibfile}

\newpage
\onecolumn
\pagenumbering{arabic}

\begin{center}
    {\Huge Supplemental Material}
\end{center}

~\\

Proofs of lemmata and propositions reported in the paper entitled ``Binary-Tree Encoding for Uniform Binary Sources in Index Modulation Systems'' are provided in the appendices below.  Additional numerical results are included for completeness at the end of this document.

\appendices

\section{Proof of Proposition~\ref{prop:complete}}\label{app:complete}
Define the mapping $\wl : \cT_{v-1} \to \cT_v$ for any integers $v > 1$ and $\ell \geq 0$, such that $\wl$ appends the protograph $\tau$ to the left-most available node at height $\ell$, relative to the deepest leaf, of a tree in $\cT_{v-1}$.  Hence, for a tree $t'\in\cT_{v-1}$ with maximum depth $d$, $\wl$ connects two edges to the left-most leaf node in level $d-\ell$ of $t'$.  These edges are, in turn, each connected to a leaf node at depth $d-\ell+1$.  Note that $w_0$ always maps a tree to a new tree with one additional internal node and one additional leaf, whereas $\wl$, for $\ell > 0$, will return the empty set if no height-$\ell$ leaves exist in the tree on which $\wl$ acts.  Algorithm~\ref{alg:badiu} applies $w_0$ and $w_1$ to each tree in $\cT_{k}$ with every step of the for loop.  The following lemma guarantees that $w_0$ and $w_1$ generate unique trees.

\begin{lem}\label{lem:1}
When applied to elements of $\cT_{v-1}$, the mappings $w_0$ and $w_1$ yield nonisomophic trees.
\end{lem}
\begin{IEEEproof}
The mapping $w_0$ is one-to-one.  Hence, the image $w_0(\cT_{v-1})$ consists of $T_{v-1}$ trees.  None of these trees are isomorphic, since no trees in $\cT_{v-1}$ are isomorphic.  Similarly, where admissible, the mapping $w_1$ is one-to-one.  In the inadmissible case where $w_1$ returns the empty set, the mapping is many-to-one; but this can be ignored, since no tree is generated. Thus, the image $w_1(\cT_{v-1})$ consists of \emph{at most} $T_{v-1}$ trees.  Again, none of these trees are isomorphic, since no trees in $\cT_{v-1}$ are isomorphic.  Furthermore, we deduce that $w_0(\cT_{v-1}) \cap w_1(\cT_{v-1}) = \{\}$, since every $t \in w_0(\cT_{v-1})$ has two leaves at the deepest level and every $t' \in w_1(\cT_{v-1})$ has more than two leaves at the deepest level.
\end{IEEEproof}

We now state the following lemma, which concludes the proof.

\begin{lem}\label{lem:2}
  For $v\geq 2$,
  \begin{equation}
    \cT_v = w_0(\cT_{v-1}) \cup w_1(\cT_{v-1})
  \end{equation}
  is a complete reduced set of $v$-node full binary trees.
\end{lem}
\begin{IEEEproof}
Let $\cT_1 = \{\tau\}$.  It is easy to verify (cf. Fig.~\ref{fig:Tv}) that $\cT_2 = w_0(\cT_1) \cup w_1(\cT_1)$ and $\cT_3 = w_0(\cT_2) \cup w_1(\cT_2)$ are complete sets.  Moreover, Lemma~\ref{lem:1} ensures $\cT_v$ contains no isomorphic trees for $v>1$.  Hence, to prove that $\cT_v$ is a complete reduced set of full binary trees for $v\geq 4$, we must show that
\begin{equation}\label{eq:setrel}
    \wl(\cT_{v-1}) \subseteq w_0(\cT_{v-1}) \cup w_1(\cT_{v-1})
\end{equation}
for $2 \leq \ell \leq v-1$ and $v\geq 4$.  

Assume the lemma is true for all $v = 2, \ldots, k-1$. Choose $t' \in \cT_{k-1}$.  Consider the mapping $\wl(t')$.  We treat several possibilities.  If the operation maps to the empty set, the set relation is satisfied since by definition (no tree is generated).  On the other hand, if $\wl(t')$ is nonempty and the deepest level of $\wl(t')$ contains exactly two leaves (and hence the same can be said for $t'$ since $\ell \geq 2$), then we must show that there exists a tree $t \in \cT_{k-1}$ such that $\wl(t') = w_0(t)$.  Note that, in this case, $w_0$ has an inverse, and the composition $w_0^{-1} \circ \wl$ commutes.  Thus, we write
\begin{equation}
  t = w_0^{-1}\circ \wl(t') = \wl \circ w_0^{-1}(t') = \wl(t'')
\end{equation}
where, in the second and third equalities, it is understood that $\wl$ operates on the level at height $\ell$ relative to the deepest leaf in $t'$.  But, by the inductive hypothesis and Lemma~\ref{lem:1}, we have that $t''=w_0^{-1}(t') \in \cT_{k-2}$.  It follows that
\begin{equation}
  t = \wl(t'') \in w_0(\cT_{k-2}) \cup w_1(\cT_{k-2}) = \cT_{k-1}
\end{equation}
as required.

Now suppose $\wl(t')$ is nonempty and the deepest level of $\wl(t')$ contains more than two leaves.  In this case, we must show that there exists a tree $t \in \cT_{k-1}$ such that $\wl(t') = w_1(t)$.  We take a similar approach, recognizing that $w_1$ has an inverse, and the composition $w_1^{-1} \circ \wl$ commutes.  It follows that
\begin{equation}
  t = w_1^{-1}\circ \wl(t') = \wl \circ w_1^{-1}(t') = \wl(t'')
\end{equation}
and, by induction, $t''=w_1^{-1}(t') \in \cT_{k-2}$.  Finally, we have that 
\begin{equation}
  t = \wl(t'') \in w_0(\cT_{k-2}) \cup w_1(\cT_{k-2}) = \cT_{k-1}
\end{equation}
as required.
\end{IEEEproof}

\section{Proof of Proposition~\ref{prop:bound}}\label{app:bound}
The proposition can be seen to hold (with equality) for $v=2,3,4$ by explicit construction of $\cT_v$.  For $v > 4$, consider the mappings $\{\wl\}$ given in Appendix~\ref{app:complete}.  As noted in the proof of Lemma~\ref{lem:1} in that appendix, $w_0$ is one-to-one.  Thus, $\vert w_0(\cT_{v-1})\vert = T_{v-1}$.  Moreover, from Lemma~\ref{lem:1} and the definition of $\cT_v$ given in Lemma~\ref{lem:2}, we know that
\begin{equation}
  T_v = \vert w_0(\cT_{v-1}) \cup w_1(\cT_{v-1}) \vert = T_{v-1} + \vert w_1(\cT_{v-1}) \vert.
\end{equation}

The set $\cT_{v-1}$ can be partitioned into two subsets: one set that contains trees that are mapped to $v$-node trees in $\cT_v$ under $w_1$ and one set does not admit a mapping under $w_1$.  We call trees in the first subset $\cT_{v-1}^o$ \emph{open} trees and trees in the second subset $\cT_{v-1}^c$ \emph{closed} trees.  Noting that $\cT_{v-1}^o = \cT_{v-1} - \cT_{v-1}^c$, we have
\begin{equation}
  T_v = T_{v-1} + \vert \cT_{v-1}^o \vert = 2 T_{v-1} - \vert \cT_{v-1}^c \vert.
\end{equation}

To lower bound $\vert \cT_{v-1}^c \vert$, we apply the following reasoning.  A $(v-1)$-node closed tree is formed by appending a closed subtree of size, say, $r$ internal nodes to a subtree of size $v-1-r$.  Each set $\cT_r$ with $r = 2^q - 1$ for some positive integer $q$ has exactly one \emph{dense} closed tree, i.e., a tree where every level is fully connected to the pervious and next levels, and the deepest level has $2^q$ leaves.  Thus, for every $q \in {2,\ldots,\floor{\lt(v-1)}}$, we can enumerate $v-1-r = v - 2^q$ closed trees.  This is a lower bound, since other combinations of $r$-node closed subtrees and $(v-1-r)$-node trees exist.  Finally, we note that if $v-1$ is one less than a power of two, $\cT_{v-1}$ contains a dense subtree, which gives rise to the $\delta_v$ parameter stated in the proposition.

\section{Proof of Lemma~\ref{lem:concave}}
\label{app:concave}
Referring to~\eqref{eq:opt_MI_p}, the equality constraint is affine and the inequality constraints are convex.  Now, consider the objective function
\begin{align}\label{eq:MI}
I(\bm{X};\bm{Y}) &= h(\bm{Y}) - h(\bm{Z}) \nonumber\\
&= -\int f_{\bm{Y}}(\mathbf{y}) \ln f_{\bm{Y}}(\mathbf{y}) \prod_{l=1}^N \operatorname{d}\!y_l\operatorname{d}\!y_l^* - n\ln(\pi e \sigma^2).
\end{align}
Hence, we must prove that $h(\bm{Y})$ is concave in $\mathbf{p}$.  Let us interpret $f_{\bm{Y}}$ (cf.~\eqref{eq:fY}) as a function of $\mathbf{p}$ for a fixed $\mathbf{y}$.  Then the mapping $f_{\bm{Y}}: [0,1]^N \to [0,\infty)$ is linear in this context.  Furthermore, $h(\bm{Y}) = \int u(f_{\bm{Y}}(\mathbf{y}))\operatorname{d}\!\mathbf{y}$, where $u(x) = -x \ln x$ is concave.  It is well known that a composition of a concave function and a linear function is concave, and that concavity is preserved under nonnegative weighted integration~\cite[sec.~3.2]{Boyd2004}.  The weight function here is simply one, so it follows that $h(\bm{Y})$ is concave in $\mathbf{p}$.

\section{Proof of Proposition~\ref{prop:awgn}}
\label{app:awgn}
Starting from Lemma~\ref{lem:concave}, we form the KKT conditions
\begin{align}\label{eq:kkt}
    -\mathbf{p}                 & \leq \bm{0} \nonumber \\
    \bm{1}^T\mathbf{p} - 1 & = 0 \nonumber \\
    \bm{\lambda_0}     & \geq \bm{0} \nonumber \\
    \operatorname{diag}(\bm{\lambda}_0)\mathbf{p} & = \bm{0} \nonumber \\
    \nabla L(\mathbf{p},\bm{\lambda}_0,\nu) &= 0
\end{align}
where $\bm{\lambda}_0 \in \mathbb{R}^C$ and $\nu \in \mathbb{R}$ are the Lagrange multipliers.  The gradient of the Lagrangian is
\begin{equation}
    \nabla L = \nabla h(\bm{Y}) + \bm{\lambda}_0  + \nu \bm{1}
\end{equation}
and the gradient of $h(\bm{Y})$ with respect to $\mathbf{p}$ can be written as
\begin{equation}
    \nabla h(\bm{Y}) = -\int \ln(\bm{a}(\mathbf{y})^T \mathbf{p})\bm{a}(\mathbf{y}) + \bm{a}(\mathbf{y}) \operatorname{d}\!\mathbf{y}
\end{equation}
where $\bm{a}(\mathbf{y}) \coloneqq (f_{\bm{Y}\mid U}(\mathbf{y}\mid U=i))_{i=1,\ldots,C}$.  Hence, the optimal vector $\mathbf{p}$ must satisfy
\begin{equation}\label{eq:kkt-int}
    \int \ln(\bm{a}(\mathbf{y})^T \mathbf{p})\bm{a}(\mathbf{y}) \operatorname{d}\!\mathbf{y} = \bm{\lambda}_0 + (\nu-1) \bm{1}.
\end{equation}
where we have used the fact that $\int \bm{a}(\mathbf{y}) \operatorname{d}\!\mathbf{y} = \bm{1}$.  Furthermore, since the problem is concave, any $\mathbf{p}$, $\bm{\lambda}_0$, and $\nu$ that satisfy~\eqref{eq:kkt-int} and the rest of the conditions in~\eqref{eq:kkt} are primal and dual optimal, and thus yield the maximum mutual information.  Choose $p_i = 1/C$ for all $i$.  In this case, the inequality constraints are inactive, which implies $\bm{\lambda}_0=\bm{0}$.  As a result,~\eqref{eq:kkt-int} is satisfied (along with the rest of the KKT conditions), and thus uniform SAP probabilities is optimal.

\section{Proof of Proposition~\ref{prop:jensen}}
\label{app:jensen}
Using Jensen's inequality and considering the concavity of the logarithm, $h(\bm{Y})$ can be lower bounded as   
\begin{align}\label{jensenpro1}
h(\bm{Y})= -\mathbb{E}\left[ \ln f_{\bm{Y}}(\mathbf{y})   \right] \geq - \ln \mathbb{E} \left[f_{\bm{Y}}(\mathbf{y})\right]   
\end{align}
where 
\begin{align}\label{jensenpro2}
\mathbb{E} \left[f_{\bm{Y}}(\mathbf{y})\right]&=\sum_{i\in \mathcal{U}}^{}\sum_{j\in \mathcal{U}}^{}p_ip_j\int_{}f_{\bm{Y}|{U}}(\mathbf{y}|U=i)f_{\bm{Y}|{U}}(\mathbf{y}|U=j)\prod_{l=1}^{N}\operatorname{d}\! y_l\operatorname{d}\! y_l^*\\ \nonumber & =\sum_{i\in \mathcal{U}}^{}\sum_{j\in \mathcal{U}}^{}\frac{p_ip_j}{\pi^{2N}\sigma^{4(N-K)}\psi_{ij}}\int_{}\prod_{l\in \mathcal{S}_i}\exp\Big(-\frac{|y_l|^2}{g_l\rho_{li}+\sigma^2}\Big)\prod_{l\notin \mathcal{S}_i}\exp\Big(-\frac{|y_l|^2}{\sigma^2}\Big)\\ \nonumber & \times \prod_{l\in \mathcal{S}_j}\exp\Big(-\frac{|y_l|^2}{g_l\rho_{lj}+\sigma^2}\Big)\prod_{l\notin \mathcal{S}_j}\exp\Big(-\frac{|y_l|^2}{\sigma^2}\Big)\prod_{l=1}^{N}\operatorname{d}\! y_l\operatorname{d}\! y_l^*\\ \nonumber &
\end{align}
and $\psi_{ij}=\prod_{l\in \mathcal{S}_i}({g_l \rho_{li}+\sigma^2})\prod_{l\in \mathcal{S}_j}({g_l \rho_{lj}+\sigma^2})$. The integrals in the expectation can be evaluated, which yields
\begin{align}\label{jensenpro3}
    \mathbb{E} \left[f_{\bm{Y}}(\mathbf{y})\right]=\sum_{i \in \mathcal{U}}\sum_{j \in \mathcal{U}}  \frac{p_i p_j}{\pi^N\det(\bm{\Xi}_i+\bm{\Xi}_j)}
\end{align}
where $\bm{\Xi}_i$ is a diagonal matrix with the $l$th element of the diagonal $\xi_{li}$ satisfying
\begin{equation}\label{jensenpro4}
    \xi_{li} = 
    \begin{cases}
    g_l \rho_{li} + \sigma^2,\quad &\text{if } l\in\cS_i, \\
    \sigma^2,&\text{otherwise.} 
    \end{cases}
\end{equation}
Substituting \eqref{jensenpro3} into \eqref{jensenpro1}, factoring out $1/\pi^N$, and using~\eqref{eq:MI_basic}, we arrive at the bound stated in the proposition.

\section{Proof of Proposition~\ref{prop:jensen-opt}}
\label{app:jensen-opt}
Let $a_{ij} = 1/\det(\bm{\Xi}_i+\bm{\Xi}_j)$.  Then Proposition~\ref{prop:jensen} can be written as
\begin{align}
    I(\mathbf{p},\bm{\rho},\sigma^2) &\geq -\ln\left(\sum_{i \in \mathcal{U}}\sum_{j \in \mathcal{U}}  a_{ij} p_i p_j\right)-N\ln\left( e \sigma^2\right) \nonumber \\
    & = -\ln(\mathbf{p}^T \bm{A} \mathbf{p}) -N\ln\left( e \sigma^2\right)
\end{align}
where $\bm{A} = (a_{ij})$.  Thus, maximizing the bound with respect to the activation probabilities (subject to constraints) can be achieved by solving the following optimization problem:
\begin{equation}
\begin{aligned}
    &\underset{\mathbf{p}}{\text{minimize}} && \mathbf{p}^T \bm{A} \mathbf{p} \\
    &\text{subject to} && \sum_{i\in\mathcal{U}} p_i = 1 \\
    &&& p_i \geq 0,\quad\forall i\in\cU.
\end{aligned}
\end{equation}
When $\bm{A}$ is nonsingular, this classical quadratic program yields the solution
\begin{equation}
    \mathbf{p}^\star = (\nu \bm{A}^{-1}\bm{1})^+
\end{equation}
where $\nu$ is chosen to satisfy the equality constraint $\sum_{i\in\mathcal{U}} p_i = 1$.  Letting $b_{ij}$ denote the $ij$th element of $\bm{A}^{-1}$, we have that $p_i^\star = \nu(\sum_{j\in\cU}b_{ij})^+$ and $\nu = 1/\sum_{i\in\cU}(\sum_{j\in\cU}b_{ij})^+$.  The result stated in the proposition follows readily.

\section{Proof of Proposition~\ref{prop:p_opt_SNRinf}}
\label{app:p_opt_SNRinf}
In the following we denote $a_{li}=g_l\rho_{li}$, for all $l$. From~\eqref{eq:MI}, together with~\eqref{eq:fY} and~\eqref{eq:fYcondX}, we have
\begin{multline}\label{eq:MI_g}
I(\mathbf{p},\bm{\rho},\sigma^2) =-\sum_{i\in\mathcal{U}}p_i\int \prod_{l\in\mathcal{S}_i} f_\text{CN}(y_l;a_{li}+\sigma^2) \prod_{l\notin\mathcal{S}_i} f_\text{CN}(y_l;\sigma^2) \\ 
\times \ln \left[ \sum_{j\in\mathcal{U}} p_j \prod_{m\in\mathcal{S}_j} f_\text{CN}(y_m;a_{mj}+\sigma^2) \prod_{m\notin\mathcal{S}_j} f_\text{CN}(y_m;\sigma^2) \right] \prod_{l=1}^N \operatorname{d}\!y_l\operatorname{d}\!y_l^* -N\ln(\pi e\sigma^2)
\end{multline}
In the integral corresponding to the $i$th term above, we make the following change of variables: $y_l\to \sqrt{a_{li}+\sigma^2} y_l$ and $y_l^*\to \sqrt{a_{li}+\sigma^2} y_l^*$, $l\in\mathcal{S}_i$, and $y_l\to\sigma y_l$ and $y_l^*\to  \sigma y_l^*$, $l\notin\mathcal{S}_i$, and obtain
\begin{multline}\label{eq:MI_ch}
I(\mathbf{p},\bm{\rho},\sigma^2) =-\sum_{i\in\mathcal{U}}p_i\int\prod_{l=1}^N \frac{e^{-|y_l|^2}}{\pi} \ln \left[ \sum_{j\in\mathcal{U}}  \frac{p_j}{(\pi\sigma^2)^{n-k}\prod_{m\in\mathcal{S}_j}\pi(a_{mj}+\sigma^2)}\right. \\ \left. \times \prod_{m\in(\mathcal{S}_j\cap\mathcal{S}_i)\cup(\mathcal{S}_j^c\cap\mathcal{S}_i^c)} e^{-|y_m|^2} \prod_{m\in\mathcal{S}_j\setminus\mathcal{S}_i} e^{-\frac{\sigma^2|y_m|^2}{a_{mj}+\sigma^2}} \prod_{m\in\mathcal{S}_i\setminus\mathcal{S}_j} e^{-\frac{(a_{mi}+\sigma^2)}{\sigma^2}|y_m|^2} \right] \prod_{l=1}^N \operatorname{d}\!y_l\operatorname{d}\!y_l^*  -N\ln(\pi e\sigma^2)
\end{multline}
In the argument of the log in~\eqref{eq:MI_ch}, we factor out $\frac{1}{(\pi\sigma^2)^N}\prod_{m=1}^N \exp\left(-|y_m|^2\right)$, such that, upon taking the log and integrating, eq.~\eqref{eq:MI_ch} becomes
\begin{multline}\label{eq:MI_dif}
I(\mathbf{p},\bm{\rho},\sigma^2) = -\sum_{i\in\mathcal{U}}p_i\int\prod_{l=1}^N \frac{e^{-|y_l|^2}}{\pi} \\ \times \ln \left[ \frac{p_i}{\prod_{m\in\mathcal{S}_i}(1+a_{mi}/\sigma^2)} + \sum_{j\in\mathcal{U}\setminus\{i\}}  \frac{p_j}{\prod_{m\in\mathcal{S}_j}(1+a_{mj}/\sigma^2)} \prod_{m\in\mathcal{S}_j\setminus\mathcal{S}_i} e^{\frac{a_{mj}|y_m|^2}{a_{mj}+\sigma^2}} \prod_{m\in\mathcal{S}_i\setminus\mathcal{S}_j} e^{-\frac{a_{mi}}{\sigma^2}|y_m|^2} \right] \prod_{l=1}^N \operatorname{d}\!y_l\operatorname{d}\!y_l^*
\end{multline}
Now, we define the probability values
\begin{equation}\label{eq:q}
q_i = \frac{\prod_{m\in\mathcal{S}_i}(a_{mj}+\sigma^2)}{\sum_{j\in\mathcal{U}} \prod_{l\in\mathcal{S}_j}(a_{lj}+\sigma^2)}
\end{equation}
and obtain
\begin{multline}\label{eq:MI_pq}
I(\mathbf{p},\bm{\rho},\sigma^2) = \ln \left[\sum_{i\in\mathcal{U}} \prod_{l\in\mathcal{S}_i}\left(\frac{a_{li}}{\sigma^2}+1\right) \right] \\
-\sum_{i\in\mathcal{U}}p_i\int\prod_{l=1}^N \frac{e^{-|y_l|^2}}{\pi} \ln \left[ \frac{p_i}{q_i} + \sum_{j\in\mathcal{U}\setminus\{i\}}  \frac{p_j}{q_j} \prod_{m\in\mathcal{S}_j\setminus\mathcal{S}_i} e^{\frac{a_{mj}|y_m|^2}{a_{mj}+\sigma^2}} \prod_{m\in\mathcal{S}_i\setminus\mathcal{S}_j} e^{-\frac{a_{mi}}{\sigma^2}|y_m|^2} \right] \prod_{l=1}^N \operatorname{d}\!y_l\operatorname{d}\!y_l^*
\end{multline}
We observe that the limit of the $i$th term of the sum in~\eqref{eq:MI_pq} as $\sigma^2\to 0$ is $p_i\ln\frac{p_i}{q_i}$ (with the convention $0\cdot\ln0=0$).\footnote{By the dominated convergence theorem, the order of the limit and the integral can be exchanged.} Thus, the mutual information obeys the asymptotic equivalence
\begin{equation}\label{eq:MI_scaling_highSNR}
I(\mathbf{p},\bm{\rho},\sigma^2) 
\sim \ln \left[\sum_{i\in\mathcal{U}} \prod_{l\in\mathcal{S}_i}\left(\frac{a_{li}}{\sigma^2}+1\right) \right] -\sum_{i\in\mathcal{U}}p_i \ln \frac{p_i}{q_i} . 
\end{equation}
Note that $\sum_{i\in\mathcal{U}}p_i \ln \frac{p_i}{q_i}$ is the Kullback-Leibler divergence between $p$ and $q$, which is always positive and equals zero if and only if the two distributions are identical. Thus, at high SNR ($\sigma^2\to 0$), the probabilities that maximize the mutual information are $p_i=q_i$, $i\in\mathcal{U}$.

\section{Proof of Corollary~\ref{cor:UB}}
\label{app:UB}
For an arbitrary probability distribution $p$, define $\mathcal{U}_+=\{i\in\mathcal{U}\mid p_i>0\}$. Based on~\eqref{eq:MI_pq}, we write
\begin{multline}\label{eq:UB}
\mu(\sigma^2) - I(\mathbf{p},\bm{\rho},\sigma^2) =  -\sum_{i\in\mathcal{U}_+}p_i \ln \frac{q_i}{p_i}\\
+\sum_{i\in\mathcal{U}_+} p_i\int \prod_{l=1}^N \frac{e^{-|y_l|^2}}{\pi} \ln \left[ 1 + \sum_{j\in\mathcal{U}\setminus\{i\}}  \frac{p_j q_i}{q_j p_i} \prod_{m\in\mathcal{S}_j\setminus\mathcal{S}_i} e^{\frac{a_{mj}|y_m|^2}{a_{mj}+\sigma^2}} \prod_{m\in\mathcal{S}_i\setminus\mathcal{S}_j} e^{-\frac{a_{mi}}{\sigma^2}|y_m|^2} \right] \prod_{l=1}^N \operatorname{d}\!y_l\operatorname{d}\!y_l^*
\end{multline}
Firstly, using Jensen's inequality, we have $-\sum_{i\in\mathcal{U}_+}p_i \ln \frac{q_i}{p_i}\geq -\ln \sum_{i\in\mathcal{U}_+}p_i \frac{q_i}{p_i}= -\ln \sum_{i\in\mathcal{U}_+}q_i\geq 0$. Secondly, the log in the integrand in~\eqref{eq:UB} is positive, as $\ln(1+t)>0$ for $t>0$); therefore, the integral is also positive. Thus, $\mu(\sigma^2)> I(\mathbf{p},\bm{\rho},\sigma^2)$, for any $\mathbf{p}$ and $\sigma^2$. Consequently, $\mu(\sigma^2)>\max_{\mathbf{p}} I(\mathbf{p},\bm{\rho},\sigma^2) = I^\star(\bm{\rho},\sigma^2)$. Finally, $\lim_{\sigma^2\to 0} [\mu(\sigma^2)-I^\star(\bm{\rho},\sigma^2)]=\lim_{\sigma^2\to 0} [\mu(\sigma^2)-I(\mathbf{q},\bm{\rho},\sigma^2)]=0$, as argued in Proposition~\ref{prop:p_opt_SNRinf}.

\section{Proof of Proposition~\ref{prop:p_opt_SNRlow}}
\label{app:p_opt_SNRlow}
Starting from~\eqref{eq:MI_dif}, we observe that, as $\sigma^2\to\infty$, the mutual information obeys the asymptotic equivalence
\begin{align*}\label{eq:MI_scaling_lowSNR}
I(\mathbf{p},\bm{\rho},\sigma^2)
&\sim -\sum_{i\in\mathcal{U}}p_i \ln \left[ \frac{p_i}{\prod_{l\in\mathcal{S}_i}\left(\frac{a_{li}}{\sigma^2}+1\right)} + \sum_{j\in\mathcal{U}\setminus\{i\}}  \frac{p_j}{\prod_{m\in\mathcal{S}_j}\left(\frac{a_{mj}}{\sigma^2}+1\right)} \right] \\
&=- \ln \left[ \sum_{i\in\mathcal{U}}  \frac{p_i}{\prod_{l\in\mathcal{S}_i}\left(\frac{a_{li}}{\sigma^2}+1\right)} \right] \\
&\leq - \ln \left[  \frac{\sum_{i\in\mathcal{U}} p_i}{\prod_{l\in\mathcal{S}_{i^\star}}\left(\frac{a_{li^\star}}{\sigma^2}+1\right)} \right] \\
&= \ln \left[ \prod_{l\in\mathcal{S}_{i^\star}}\left(\frac{a_{li^\star}}{\sigma^2}+1\right)\right] = \sum_{l\in\mathcal{S}_{i^\star}}\ln\left(\frac{a_{li^\star}}{\sigma^2}+1\right)
\end{align*}
The inequality is achieved when $p_i=r_i$, $i\in\mathcal{U}$.

\section{Proof of Proposition~\ref{prop:rho_opt_SNRinf_SNRlow}}
\label{app:rho_opt_SNRinf_SNRlow}
Consider the high-SNR regime.  According to Proposition~\ref{prop:p_opt_SNRinf} and Corollary~\ref{cor:UB}, the optimal mutual information for given powers satisfies the asymptotic equivalence
\begin{equation}
I^\star(\bm{\rho},\sigma^2) \sim\ln \left[\sum_{i\in\mathcal{U}} \prod_{l\in\mathcal{S}_i}\left(\frac{g_l\rho_{li}}{\sigma^2}+1\right) \right], \quad \text{as }\sigma^2\to 0.
\end{equation}
We now maximize the mutual information over the powers at high SNR by formulating the optimization problem
\begin{equation}\label{eq:opt_powers}
\begin{aligned}
    &\underset{\bm{\rho}}{\text{maximize}} && \ln \left[\sum_{i\in\mathcal{U}} \exp\sum_{l\in\mathcal{S}_i}\ln\left(\frac{g_l\rho_{li}}{\sigma^2}+1\right) \right] \\
    &&& \sum_{l\in\mathcal{S}_i}\rho_{li} \leq P \text{ and } \rho_{li}\geq 0,\quad\forall i\in\mathcal{U}, l\in\mathcal{S}_i.
\end{aligned}
\end{equation}
Given that the objective function is increasing in every variable, the constraints are satisfied with equality. Moreover, the problem actually decouples in ${N\choose K}$ separate problems. We cast the $i$th problem as
\begin{equation}
\begin{aligned}
    &\underset{\bm{\rho}}{\text{maximize}} &&  \sum_{l\in\mathcal{S}_i}\ln\left(\frac{g_l\rho_{li}}{\sigma^2}+1\right)  \\
    &&& \sum_{l\in\mathcal{S}_i}\rho_{li} \leq P \text{ and } \rho_{li}\geq 0.
\end{aligned}
\end{equation}
For each $i\in\mathcal{U}$, the optimal powers are found via the waterfilling strategy.

Now consider the low-SNR regime.  According to Proposition~\ref{prop:p_opt_SNRlow}, we optimize $I^\star(\bm{\rho},\sigma^2)\sim \sum_{l\in\mathcal{S}_{i^\star}}\ln\left(\frac{a_{li^\star}}{\sigma^2}+1\right)$ under the constraint $\sum_{l\in\mathcal{S}_i^\star}\rho_{li^\star} = P$. The optimum is given by waterfilling power allocation.

\section{Additional Numerical Results}
\label{app:results}
\begin{figure*}[!h]
    \centering
    \begin{subfigure}[t]{0.5\textwidth}
        \centering
        \includegraphics[width=8cm]{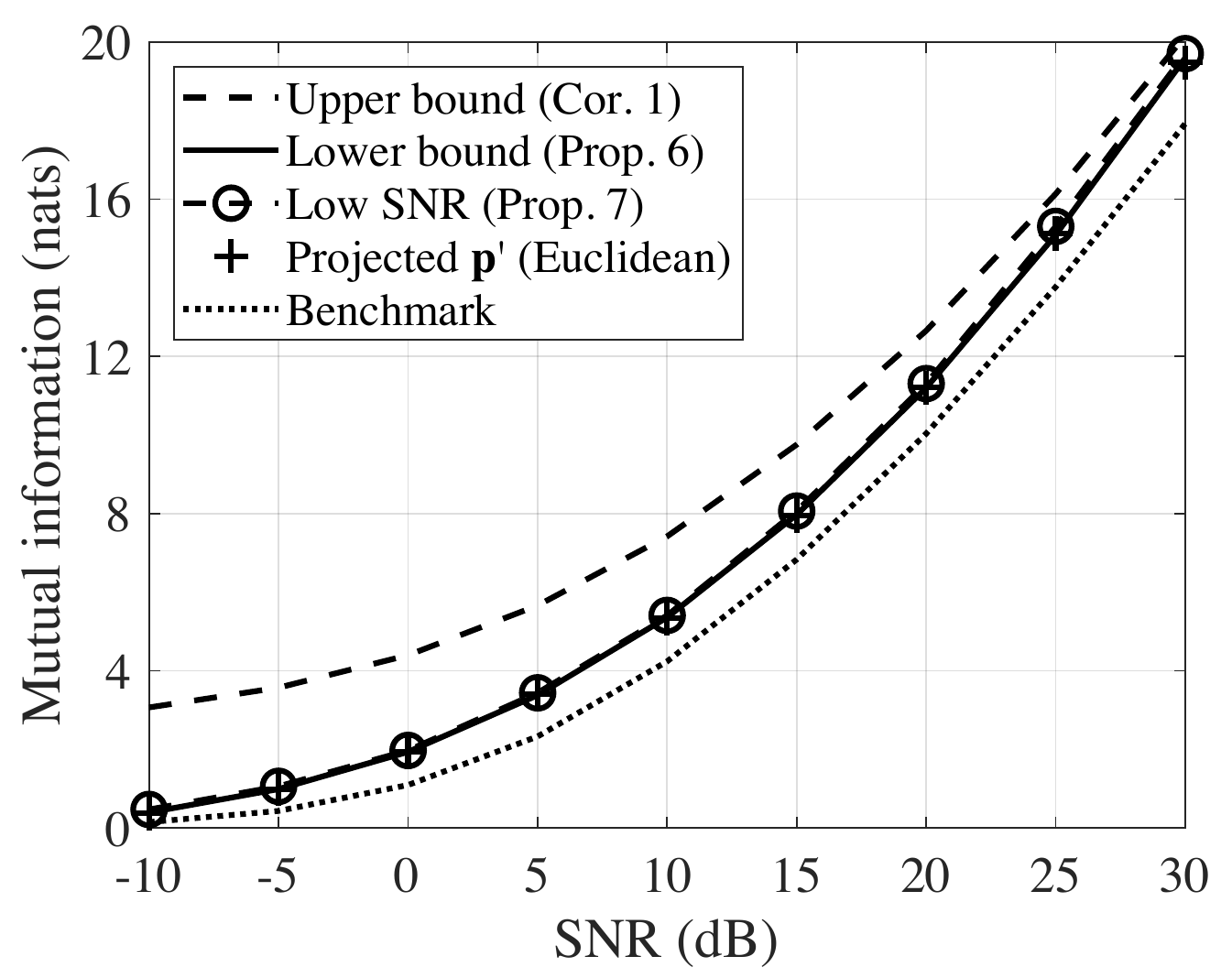}
        \caption{$\eta=0.2$}
        \label{fig:etap2_6c4}
    \end{subfigure}%
~
    \begin{subfigure}[t]{0.5\textwidth}
        \centering
        \includegraphics[width=8cm]{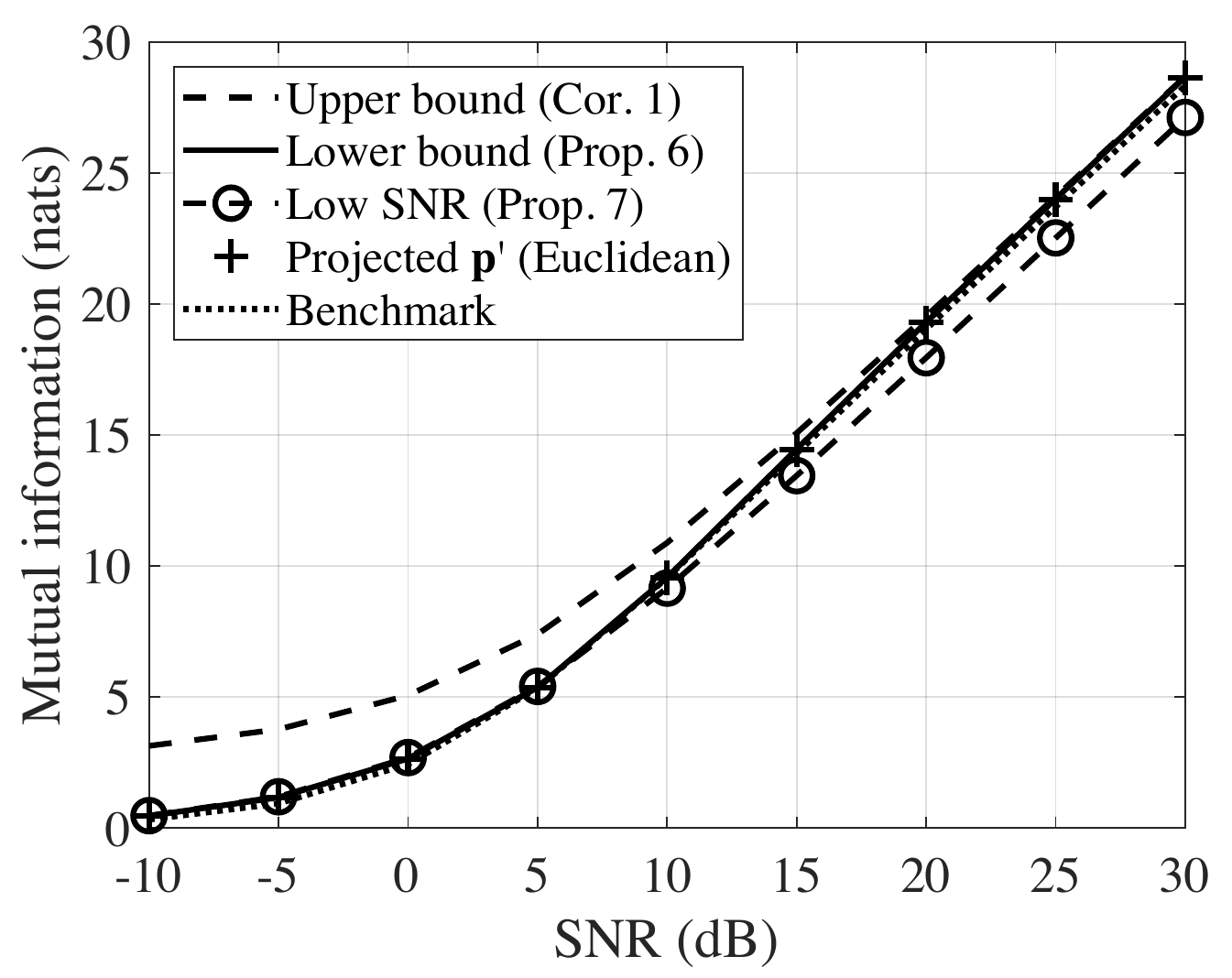}
        \caption{$\eta=0.7$}
        \label{fig:etap7_6c4}
    \end{subfigure}
    \caption{Mutual information vs. SNR for $(N,K)=(6,4)$ with $\eta=0.2$ and $\eta=0.7$, where $g_l=\eta^{l-1}$ for $l = 1,\ldots,N$.  Waterfilling power allocation is used for all results except for the benchmark, which uses uniform power allocation.  The curve labelled ``Projected $\bp'$ (Euclidean)'' gives an achievable rate when all SAPs are allowed to be utilized and power allocation is employed.}
    \label{fig:eta_6c4}
\end{figure*}

\begin{figure}[!h]
    \centering
    \includegraphics[width=8cm]{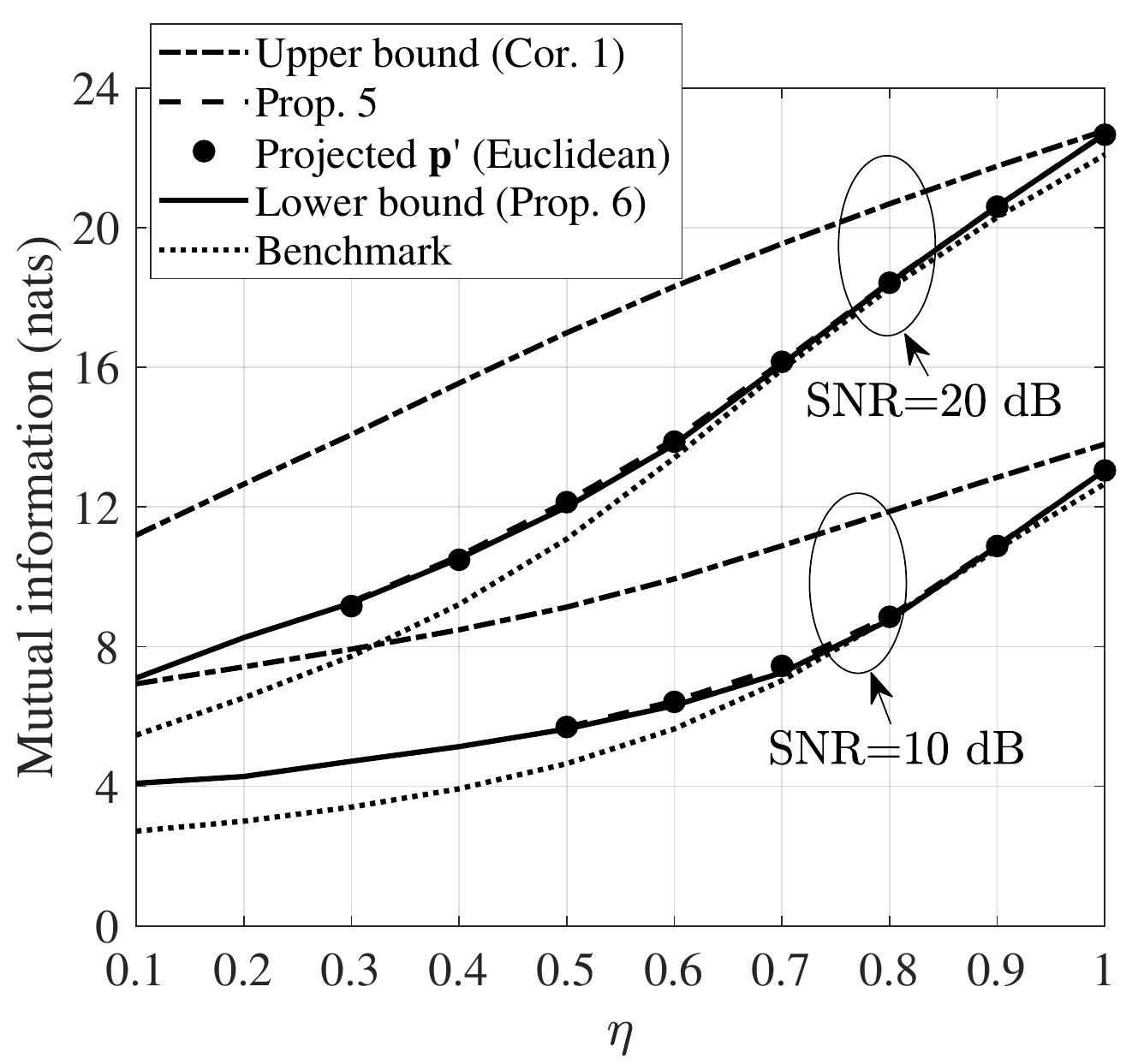}
    \caption{Mutual information vs. $\eta$ for $(N,K)=(6,4)$, where $g_l=\eta^{l-1}$ for $l = 1,\ldots,N$.  The curve labelled ``Projected $\bp'$ (Euclidean)'' gives an achievable rate when all SAPs are allowed to be utilized and power allocation is employed.}  
    \label{fig:var-eta_6c4}
\end{figure}

Here, we include additional numerical results obtained for the benefit of illustration.  Figs.~\ref{fig:eta_6c4} and~\ref{fig:var-eta_6c4} relate to systems with $N=6$ and $K=4$.  In Fig.~\ref{fig:etap2_6c4}, a clear and fairly constant benefit can be observed in highly frequency-selective channels for the optimized system compared to the benchmark OFDM-IM system, which utilizes eight SAPs out of fifteen.  The advantage is severly diminished when the channel is less selective (cf.~Fig.~\ref{fig:etap7_6c4}).  This behavior is observed more clearly in Fig.~\ref{fig:var-eta_6c4}.  Note that, in this figure, results related to the use of the probability computation detailed in Proposition~\ref{prop:jensen-opt} are not available for low values of $\eta$, because the matrix $\bm{A}$ is singular for these parameterizations.  One could use a numerical approach or one of the other analytic results detailed in Section~\ref{sec:rate-relaxed} to obtain the relaxed SAP probability distribution in this case.

\begin{figure*}[!h]
    \centering
    \begin{subfigure}[t]{0.5\textwidth}
        \centering
        \includegraphics[width=8cm]{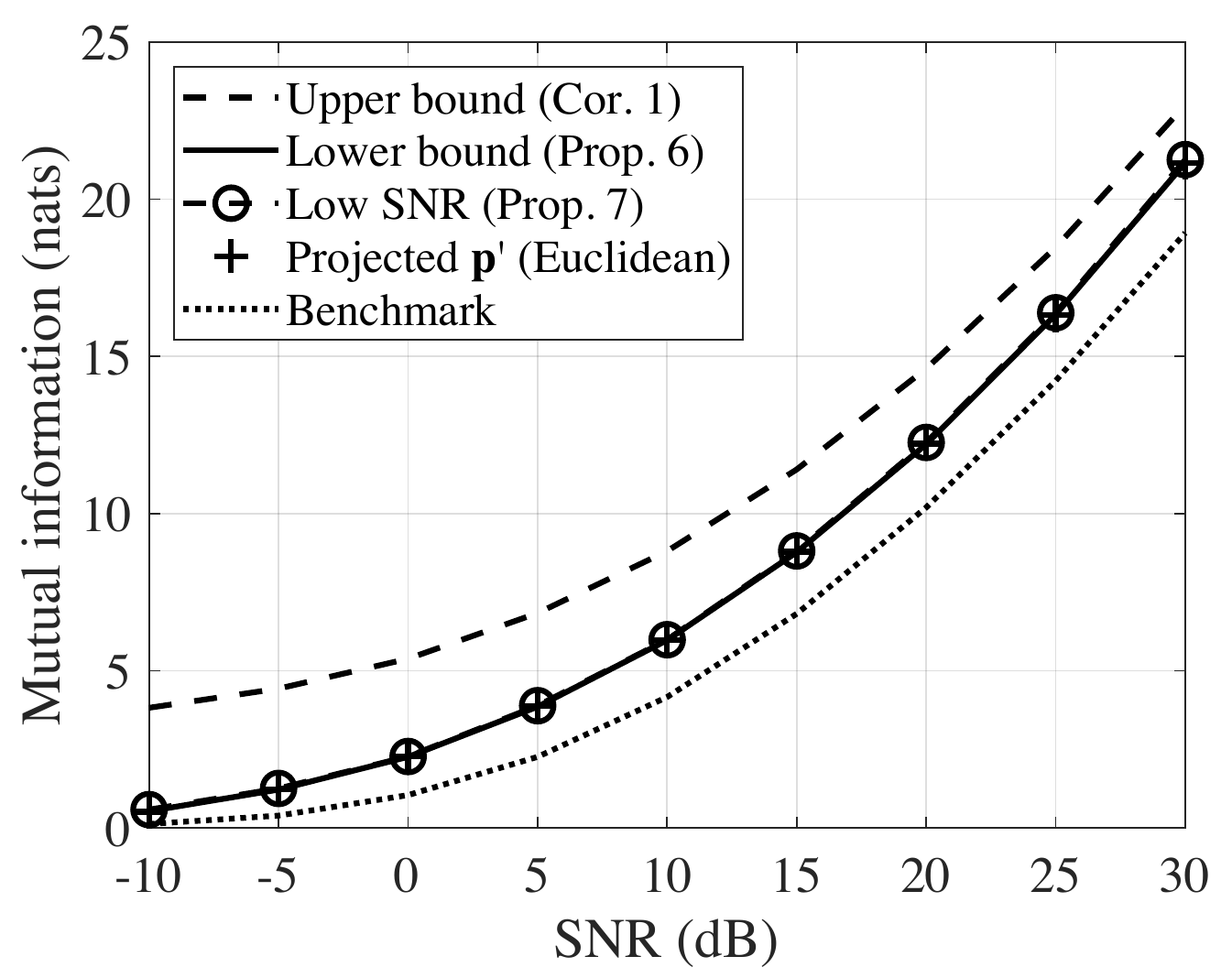}
        \caption{$\eta=0.2$}
        \label{fig:etap2_8c6}
    \end{subfigure}%
~
    \begin{subfigure}[t]{0.5\textwidth}
        \centering
        \includegraphics[width=8cm]{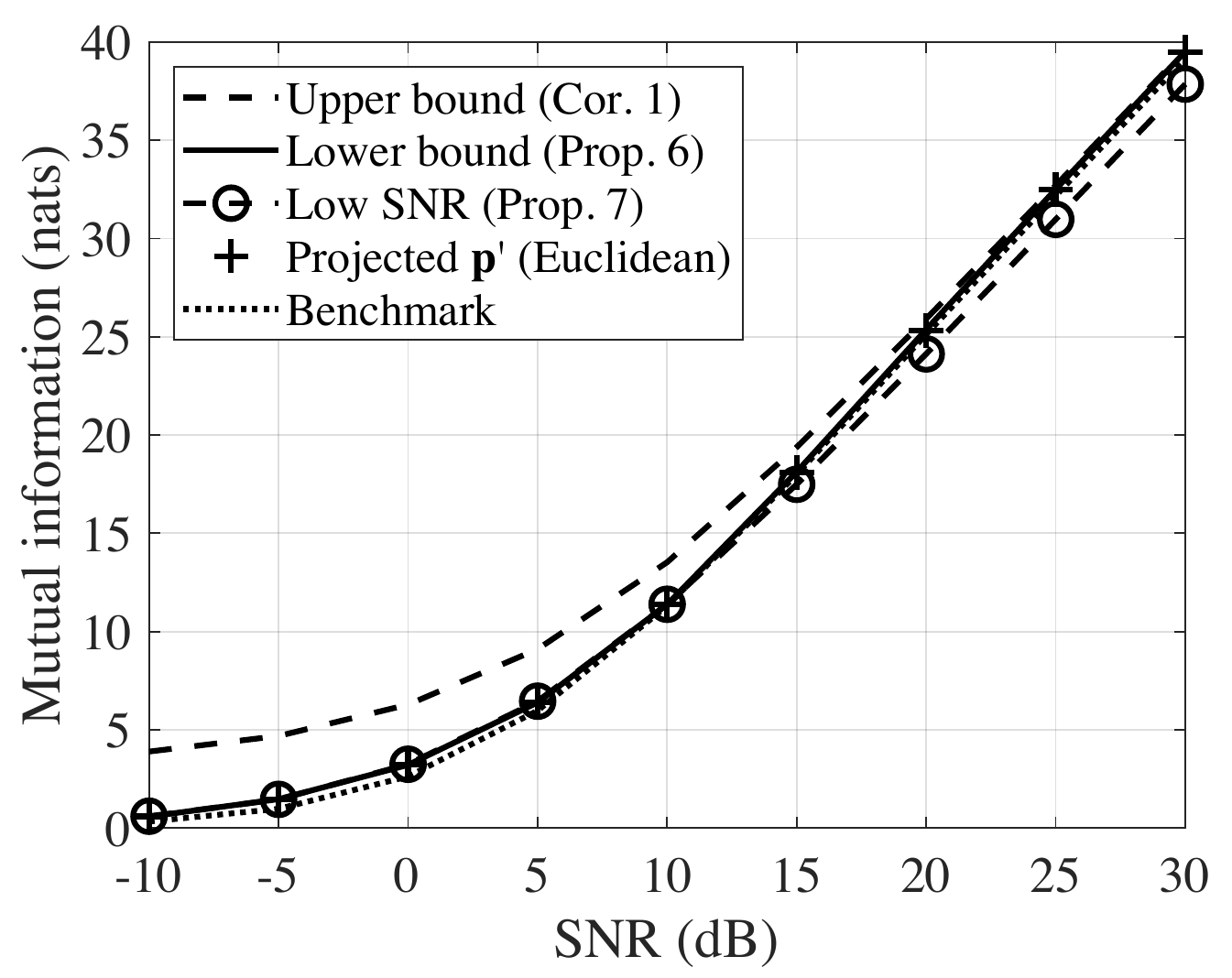}
        \caption{$\eta=0.7$}
        \label{fig:etap7_8c6}
    \end{subfigure}
    \caption{Mutual information vs. SNR for $(N,K)=(8,6)$ with $\eta=0.2$ and $\eta=0.7$, where $g_l=\eta^{l-1}$ for $l = 1,\ldots,N$.  Waterfilling power allocation is used for all results except for the benchmark, which uses uniform power allocation.  The curve labelled ``Projected $\bp'$ (Euclidean)'' gives an achievable rate when all SAPs are allowed to be utilized and power allocation is employed.}
    \label{fig:eta_8c6}
\end{figure*}

\begin{figure}[!h]
    \centering
    \includegraphics[width=8cm]{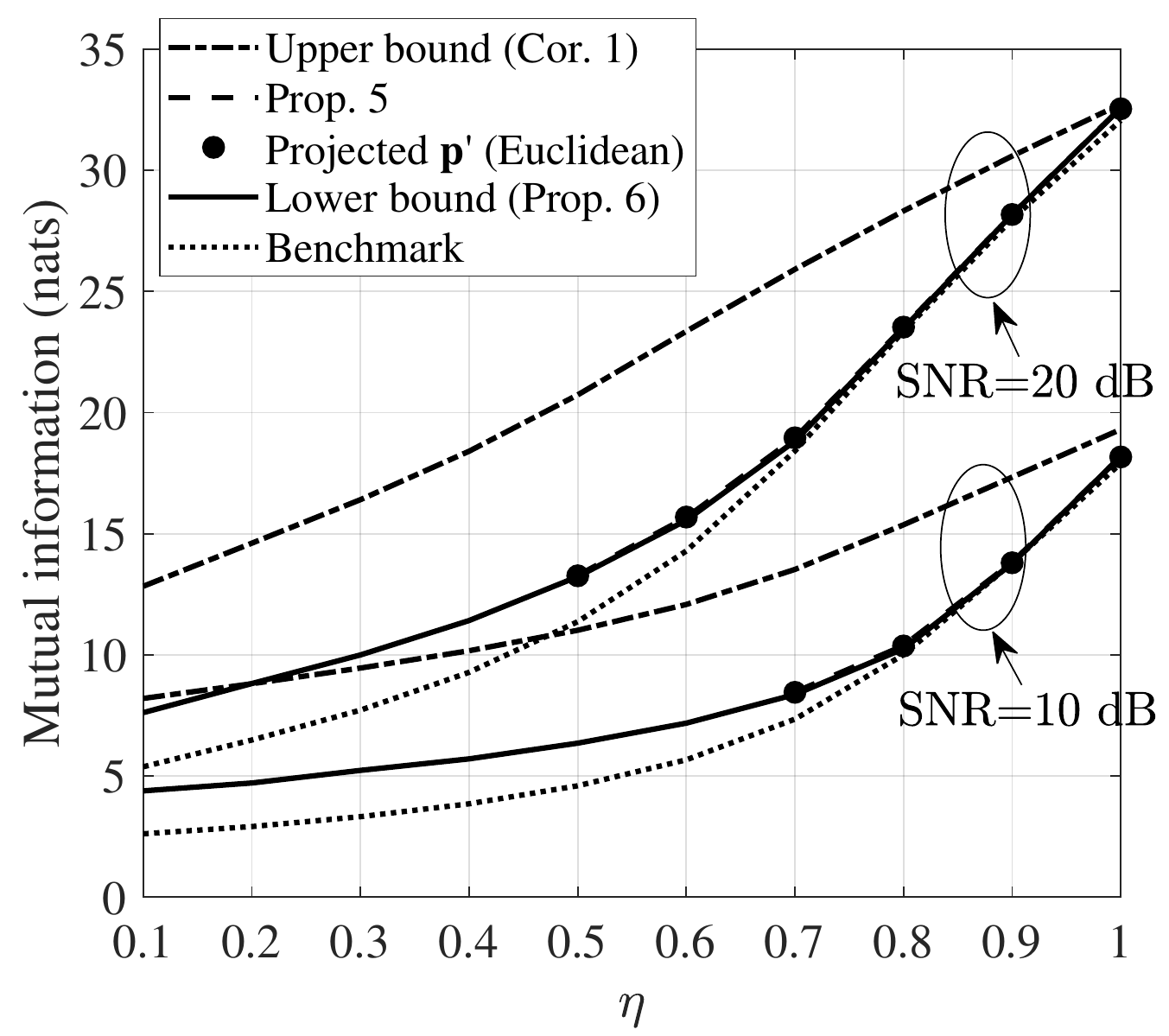}
    \caption{Mutual information vs. $\eta$ for $(N,K)=(8,6)$, where $g_l=\eta^{l-1}$ for $l = 1,\ldots,N$.  The curve labelled ``Projected $\bp'$ (Euclidean)'' gives an achievable rate when all SAPs are allowed to be utilized and power allocation is employed.}  
    \label{fig:var-eta_8c6}
\end{figure}

Figs.~\ref{fig:eta_8c6} and~\ref{fig:var-eta_8c6} relate to systems with $N=8$ and $K=6$.  In this scenario, the benchmark scheme utilizes sixteen out of twenty-eight SAPs.  As a result, the benefit offered by the optimized technique is more clearly observed for $\eta = 0.2$.  For the less frequency-selective case (Fig.~\ref{fig:etap7_8c6}), the gains, again, diminish.  In Fig.~\ref{fig:var-eta_8c6}, we see that the utility of Proposition~\ref{prop:jensen-opt} is restricted to reasonably high $\eta$ values (less frequency-selective channels).  

\end{document}